%% file: HIG-14-027_temp.tex
\pdfoutput=1

\documentclass[11pt,twoside,a4paper,cmspaper,final,collab]{cms-tdr}
\def\svnVersion{366745}\def\cmsCernNoTag{CERN-PH-EP/2015-255}\def\cmsCernDate{\today}\def\cmsMessage{Published in the Journal of High Energy Physics as \href{http://dx.doi.org/10.1007/JHEP06(2016)177}{\doi{10.1007/JHEP06(2016)177.}}}

\begin{document}\cmsNoteHeader{HIG-14-027}

\hyphenation{had-ron-i-za-tion}
\hyphenation{cal-or-i-me-ter}
\hyphenation{de-vices}
\RCS$Revision: 351131 $
\RCS$HeadURL: svn+ssh://svn.cern.ch/reps/tdr2/papers/HIG-14-027/trunk/HIG-14-027.tex $
\RCS$Id: HIG-14-027.tex 351131 2016-06-29 08:08:54Z stiegerb $
\newlength\cmsFigWidth
\ifthenelse{\boolean{cms@external}}{\setlength\cmsFigWidth{0.85\columnwidth}}{\setlength\cmsFigWidth{0.4\textwidth}}
\ifthenelse{\boolean{cms@external}}{\providecommand{\cmsLeft}{top\xspace}}{\providecommand{\cmsLeft}{left\xspace}}
\ifthenelse{\boolean{cms@external}}{\providecommand{\cmsRight}{bottom\xspace}}{\providecommand{\cmsRight}{right\xspace}}
\cmsNoteHeader{HIG-14-027}
\title{Search for the associated production of a Higgs boson with a single top quark in proton-proton collisions at $\sqrt{s} = 8\TeV$}

\newcommand{\tauh}{\ensuremath{\PGt_\mathrm{h}}\xspace}
\newcommand{\mgg}{\ensuremath{m_{\gamma\gamma}}\xspace}
\newcommand{\mh}{\ensuremath{m_{\PH}}\xspace}
\newcommand{\tth}{\ensuremath{\ttbar\PH}\xspace}
\newcommand{\thq}{\ensuremath{\PQt\PH\PQq}\xspace}
\newcommand{\ttgg}{\ensuremath{\ttbar\gamma\gamma}\xspace}
\newcommand{\tgg}{\ensuremath{\PQt\gamma\gamma}\xspace}
\newcommand{\W}{\ensuremath{\cmsSymbolFace{W}}\xspace}
\newcommand{\ZZ}{\ensuremath{\cmsSymbolFace{ZZ}}\xspace}
\newcommand{\WW}{\ensuremath{\cmsSymbolFace{WW}}\xspace}
\newcommand{\WZ}{\ensuremath{\cmsSymbolFace{WZ}}\xspace}
\newcommand{\VVV}{\ensuremath{\cmsSymbolFace{VVV}}\xspace}
\newcommand{\WVV}{\ensuremath{\cmsSymbolFace{WVV}}\xspace}
\newcommand{\ttW}{\ensuremath{\ttbar\cmsSymbolFace{W}^{\pm}}\xspace}
\newcommand{\ttWW}{\ensuremath{\ttbar\WW}\xspace}
\newcommand{\ttZ}{\ensuremath{\ttbar\cmsSymbolFace{Z}}\xspace}
\newcommand{\ttH}{\ensuremath{\ttbar\cmsSymbolFace{H}}\xspace}
\newcommand{\ttG}{\ensuremath{\ttbar\gamma}\xspace}
\newcommand{\ttV}{\ensuremath{\ttbar\mathrm{V}}\xspace}
\newcommand{\tbZ}{\ensuremath{\PQt\PQb\cmsSymbolFace{Z}}\xspace}
\newcommand{\ttGStar}{\ensuremath{\ttbar\gamma^*}\xspace}
\newcommand{\tZq}{\ensuremath{\PQt\cmsSymbolFace{Z}\PQq}\xspace}
\newcommand{\WWqq}{\ensuremath{\mathrm{W^{\pm}W^{\pm}\PQq\PQq}}\xspace}
\newcommand{\Ct}{\ensuremath{C_{\PQt}}\xspace}
\newcommand{\Hbb}{\ensuremath{\PH\to\bbbar}\xspace}
\newcommand{\HWW}{\ensuremath{\PH\to\W\W}\xspace}
\newcommand{\Hgg}{\ensuremath{\PH\to\PGg\PGg}\xspace}
\newcommand{\tHqWW}{\ensuremath{\PQt\PH(\W\W)\PQq}\xspace}
\newcommand{\tHqtt}{\ensuremath{\PQt\PH(\PGt\PGt)\PQq}\xspace}
\newcommand{\tHWWW}{\ensuremath{\PQt\PH(\W\W)\W}\xspace}
\newcommand{\tHWtt}{\ensuremath{\PQt\PH(\PGt\PGt)\W}\xspace}
\newcommand{\tHq}{\ensuremath{\PQt\PH\PQq}\xspace}
\newcommand{\tHW}{\ensuremath{\PQt\PH\W}\xspace}
\newcommand{\emu}{\ensuremath{\Pe\Pgm}\xspace}
\newcommand{\mumu}{\ensuremath{\Pgm\Pgm}\xspace}
\newcommand{\emt}{\ensuremath{\Pe\Pgm\tauh}\xspace}
\newcommand{\mmt}{\ensuremath{\Pgm\Pgm\tauh}\xspace}

\date{\today}

\abstract{
This paper presents the search for the production of a Higgs boson in
association with a single top quark (tHq), using data collected in proton-proton
collisions at a center-of-mass energy of 8\TeV corresponding to an
integrated luminosity of 19.7\fbinv.  The search exploits a variety of
Higgs boson decay modes resulting in final states with photons, bottom
quarks, and multiple charged leptons, including tau leptons, and employs a
variety of multivariate techniques to maximize sensitivity to the signal.
The analysis is optimized for the opposite sign of the Yukawa coupling
to that in the standard model, corresponding to a large enhancement of
the signal cross section. In the absence of an excess of candidate signal
events over the background predictions, 95\% confidence level observed
(expected) upper limits on anomalous tHq production are set, ranging
between 600 (450)\unit{fb} and 1000 (700)\unit{fb} depending on the assumed diphoton
branching fraction of the Higgs boson. This is the first time that results
on anomalous tHq production have been reported.}

\hypersetup{%
pdfauthor={CMS Collaboration},%
pdftitle={Search for the associated production of a Higgs boson with a single
top quark in proton-proton collisions at sqrt(s) = 8 TeV},%
pdfsubject={CMS},%
pdfkeywords={Higgs, top, Yukawa, CMS}}

\maketitle

\section{Introduction}
\label{sec:intro}

The discovery of a Higgs boson by the ATLAS and CMS experiments in 2012~\cite{Chatrchyan:2012ufa,Aad:2012tfa,Chatrchyan:2013lba} opened a new field for exploration in particle physics.
The Higgs boson was discovered through its direct coupling to other known heavy bosons (W, Z) and its indirect coupling to photons, which in the standard model (SM) occurs via a loop involving \PW\ bosons or top quarks.
Strong evidence for the Higgs boson coupling to fermions has also been established~\cite{Chatrchyan:2014vua,ATLAS:tautau}.
Moreover, there is evidence of the Higgs boson coupling to bottom quarks from the Tevatron~\cite{Aaltonen:2012qt} and from CMS~\cite{Chatrchyan:2013zna}, and to tau leptons from ATLAS~\cite{ATLAS:tautau}  and CMS~\cite{Chatrchyan:2014nva}.
It is now critical to test whether the observed Higgs boson is the SM Higgs boson by studying its coupling to other elementary particles.

\begin{figure}[th!]
        \centering
        \includegraphics[width=0.3\textwidth]{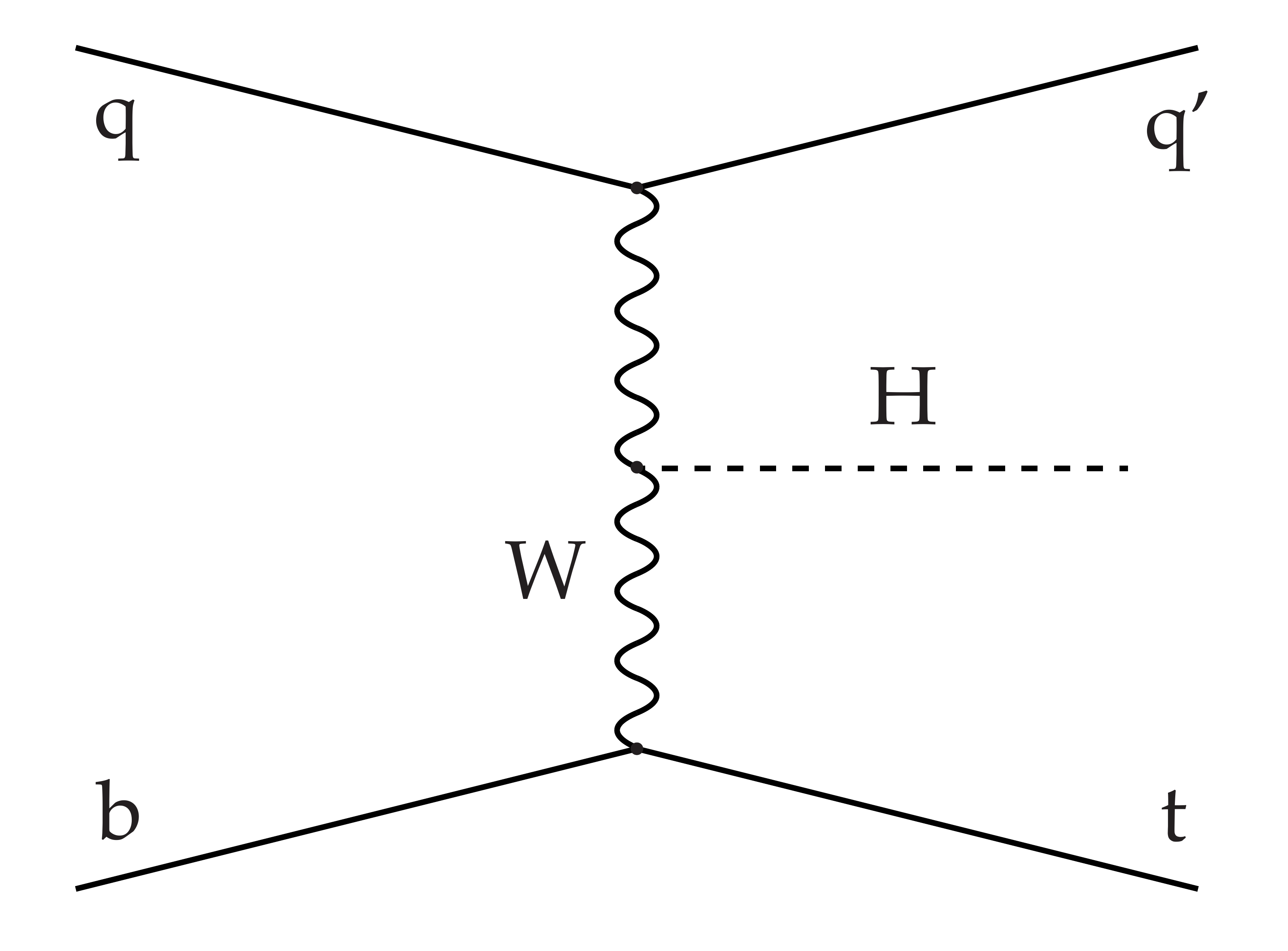}\hspace{1cm}
        \includegraphics[width=0.3\textwidth]{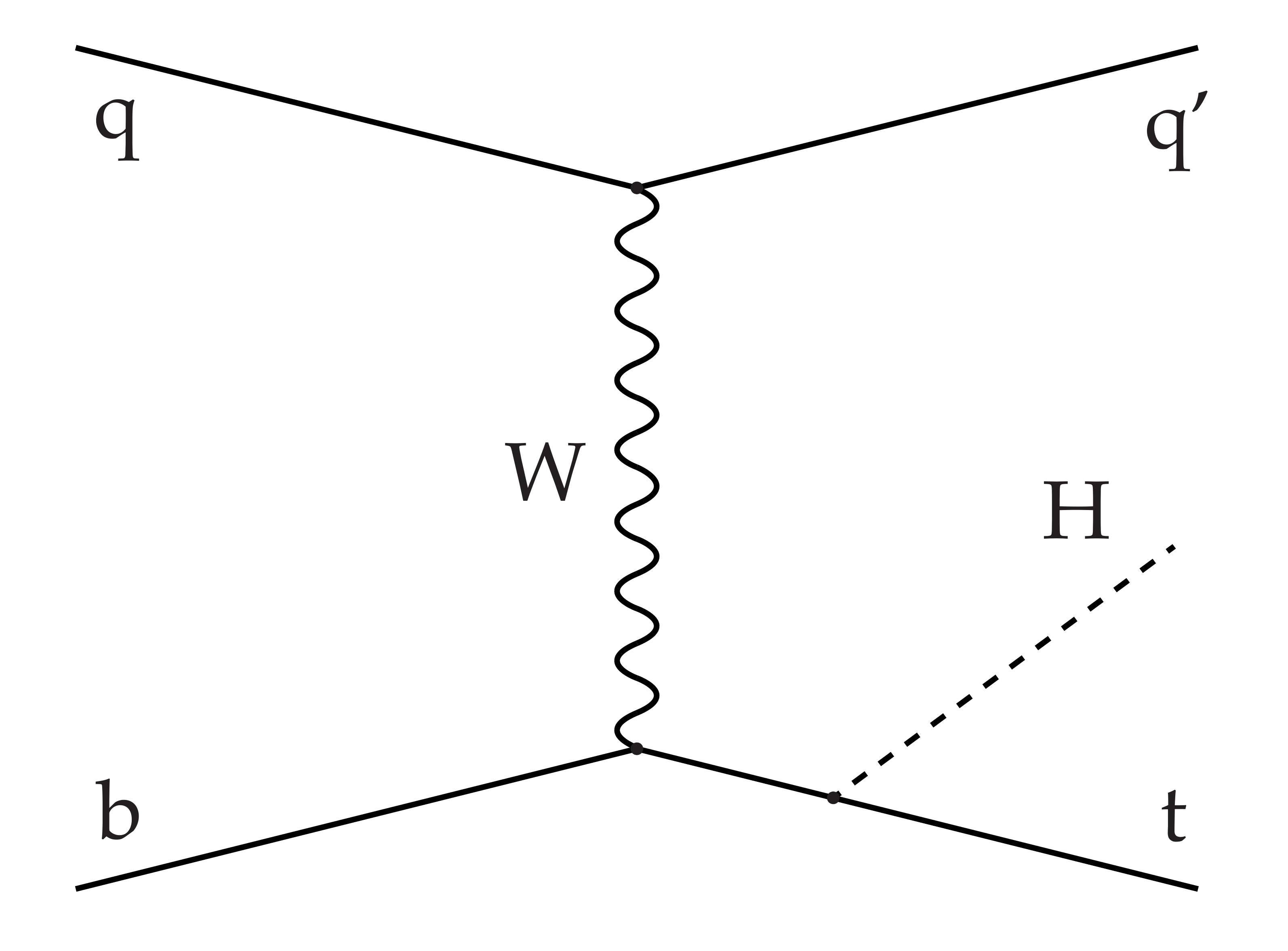}
        \caption{Dominant Feynman diagrams for the production of \tHq events: the Higgs boson is typically radiated
        from the heavier particles of the diagram,  \ie the W boson (left) or the top quark (right).\label{fig:thq_prod}}
\end{figure}

The coupling of the new boson to the top quark is of special interest. Because of its very large mass~\cite{ATLAS:2014wva} the top quark is widely believed to play a special role in the mechanism of electroweak symmetry breaking.
Physics beyond the SM could modify the top quark Yukawa coupling without violating current experimental constraints.  The most straightforward way to study this coupling is through the measurement of top quark-antiquark pair production in association with a Higgs boson (\ttH), as was recently done by ATLAS and CMS~\cite{CMS:tthcombo_paper, Khachatryan:2015ila, Aad:2014lma, Aad:2015gra}.
Interactions of the Higgs boson with the top quark can also be probed by studying the associated production of a single top quark and a Higgs boson, which proceeds mainly through $t$-channel diagrams (\tHq)~\cite{Maltoni:2001hu} in which the Higgs boson is emitted either from an internally exchanged \PW\ boson or from a top quark, as shown in Fig.~\ref{fig:thq_prod}.
The associated single top quark and Higgs boson production can also be accompanied by
a \PW\ in the final state (\tHW).
As the couplings of the Higgs boson to the W boson and the top quark have opposite signs in the SM, these two diagrams interfere destructively.  The cross section for single top quark plus Higgs boson production via the \tHq\ process in pp collisions at a center-of-mass energy of 8\TeV has been calculated to be about 18~fb at next-to-leading-order (NLO)~\cite{ennio}.

Anomalous coupling of the Higgs boson to SM particles would modify the expected rate of \tHq events~\cite{Agrawal:2012ga}. A number of models have been proposed that would modify the
interference  between the diagrams involving $\ttbar\PH$ and
$\PW\PW\PH$ couplings.
For example, a negative coupling of the Higgs boson to the top
quark~($\Ct=-1$) would give rise to about a 15-fold increase in the \tHq cross section.  Recent work suggests the
investigation of anomalous \tHq\ production in events with a pair of
photons~\cite{Biswas:2012bd,Biswas:2013xva}, b quarks~\cite{ennio}, or
multiple leptons in the final state~\cite{Biswas:2013xva}. The same
interference probes the CP-violating phase of the top quark Yukawa
coupling~\cite{Yue:2014tya,Kobakhidze:2014gqa,Demartin:2015uha}.  Also, a large rate of
single top quark plus Higgs boson events could signal the direct production
of heavy new particles as predicted in composite and little Higgs
models~\cite{Aguilar-Saavedra:2013qpa}, or new physics showing up as
Higgs boson mediated flavor changing neutral currents~\cite{Greljo:2014dka}.
 The apparent exclusion of the $\Ct=-1$ case based on the value of the branching fraction for $\PH\to\Pgg\Pgg$ only holds under the assumption that no new particles contribute to the loop in the main diagram for that decay~\cite{Ellis:2013lra}.

This paper reports the first search for \tHq\ production, focusing on the scenario where the coupling of the Higgs boson to the top quark has
a sign opposite to that predicted by the SM, using data collected by the CMS experiment at the CERN LHC. Four Higgs boson decay modes are explored.  Section~\ref{sec:exp_setup} describes the CMS detector, the reconstruction algorithms, and the
simulated samples. Section~\ref{sec:an_des} outlines the selection, background modeling, and signal extraction techniques for analyses based on \PH\ decay channels with photons, hadrons, and multiple leptons.
Section~\ref{subsec:common_systs} describes the systematic uncertainties affecting the search results.
Finally, the procedure for combining the results of the searches is presented in Section~\ref{sec:res}.
The results are summarized in Section~\ref{sec:summ}.

\section{The CMS detector, event reconstruction, and simulation}
\label{sec:exp_setup}
The central feature of the CMS apparatus is a superconducting solenoid
of 6\unit{m} internal diameter, providing an axial magnetic field of
3.8\unit{T}.
Within the magnet volume, there are a silicon pixel and strip tracker, a lead
tungstate crystal electromagnetic calorimeter (ECAL), and a brass and scintillator hadron calorimeter (HCAL), each composed of a barrel and two endcap sections.
The tracking detectors provide coverage for charged particles within pseudorapidity $\abs{\eta} < 2.5$. The ECAL
and HCAL calorimeters provide coverage up to $\abs{\eta} < 3.0$. The ECAL
is divided into two distinct regions: the barrel region, which covers
$\abs{\eta} < 1.48$, and the endcap region, which covers $1.48 < \abs{\eta} <
3.00$.  A quartz-fiber forward calorimeter extends the coverage further
up to $\abs{\eta} < 5.0$.  Muons are measured in gas-ionization detectors
embedded in the steel flux-return yoke outside the solenoid.
A detailed description of the CMS detector, together with a definition
of the coordinate system used and the relevant kinematic variables, can be
found in Ref.~\cite{Chatrchyan:2008aa}.

The particle-flow (PF) event reconstruction algorithm~\cite{CMS-PAS-PFT-09-001, CMS-PAS-PFT-10-001} consists of reconstructing and identifying each single particle with an optimized combination of all subdetector information.
The energy of charged hadrons is determined from a combination of their momentum measured in the tracker and the matching ECAL and HCAL energy deposits.

Photon PF candidates are reconstructed from the energy deposits in the ECAL, grouping the individual clusters into a supercluster.  The superclustering algorithms achieve an almost complete reconstruction of the energy of photons (and electrons) that convert into electron-positron pairs (emit bremsstrahlung) in the material in front of the ECAL.\@
The photon candidates are identified within the ECAL fiducial region $\abs{\eta} < 2.5$, excluding
the barrel-endcap transition region $1.44 < \abs{\eta} < 1.57$, where
photon reconstruction is sub-optimal.  Isolation requirements are
applied to photon candidates by looking at neighboring particle
candidates.
In the barrel section of the ECAL, an energy resolution of about 1\% is achieved for unconverted or late-converting photons in the tens of \GeVns energy range. The remaining barrel photons have a resolution of about 1.3\% up to a pseudorapidity of $\abs{\eta} = 1$, rising to about 2.5\% at $\abs{\eta} = 1.4$. In the endcaps, the resolution of unconverted or late-converting photons is about 2.5\%, while the remaining endcap photons have a resolution between 3 and 4\%.  Additional details on photon reconstruction and identification can be found in Refs.~\cite{Khachatryan:2014ira,Khachatryan:2015iwa}.

Electrons with \pt greater than 7\GeV are reconstructed within the geometrical acceptance of the tracker, $\abs{\eta} < 2.5$.
The electron momentum is determined from the combination of ECAL and tracker measurements.
Electron identification relies on a multivariate (MVA) technique, which combines observables sensitive to the amount of bremsstrahlung along the electron trajectory, the spatial and momentum matching between the electron trajectory and associated clusters, and shower shape observables~\cite{Khachatryan:2015hwa,Khachatryan:2015iwa}.
In order to increase the lepton efficiency, the $\PH\to\text{leptons}$ analysis uses a looser selection for the MVA discriminant than do the other analysis channels.

Muons with $\pt > 5\GeV$ are reconstructed within $\abs{\eta} < 2.4$~\cite{Chatrchyan:2012xi}.
The reconstruction combines information from both the silicon tracker and the muon spectrometer.
The PF muons are selected from the reconstructed muon track candidates by applying minimal
requirements on the track components in the muon and tracker systems and taking into account
matching with energy deposits in the calorimeters~\cite{CMS-PAS-PFT-10-003}.

Particles reconstructed with the PF algorithm are clustered into jets using the anti-$\kt$ algorithm with a distance parameter of 0.5~\cite{Cacciari:2008gp,Cacciari:2011ma}.
Jet energy corrections are applied to account for the non-linear response of the calorimeters to the particle energies and other detector effects.
These include corrections due to additional interactions within a beam crossing (pileup), where the average energy density from the extra interactions is evaluated on an event-by-event basis and the corresponding energy is subtracted from each jet~\cite{Cacciari:2008gn}.
The jet energy resolution is also modified in simulation with a smearing technique to match what is measured in data~\cite{Chatrchyan:2011ds}.
In all the final states that are studied, jets with $\abs{\eta} < 5.0$ and transverse momentum down to $20\GeV$ are considered, though the final selection depends on the specific analysis.

The hadronic decay of a $\Pgt$ lepton ($\tauh$) produces a narrow
jet of charged and neutral hadrons, which are mostly pions. Each neutral
pion subsequently decays into a pair of photons. The identification of
$\tauh$ jets begins with the formation of PF jets by clustering
charged hadron and photon objects via the anti-$\kt$ algorithm.
Additional details on \Pgt\ reconstruction and identification can be found in Ref.~\cite{CMS-PAS-TAU-11-001}.
For this analysis, decays involving one or three charged hadrons are used.

The missing transverse momentum vector $\ptvecmiss$ is defined as the negative projection on the plane perpendicular to the beams of the vectoral sum of the momenta of all reconstructed PF candidates in an event.
Its magnitude is referred to as \ETmiss.

Jets are identified as originating from \cPqb{}~quark production (\cPqb\ tagged) using an algorithm based on the combined properties of secondary vertices and track-based lifetime information, known as the combined secondary vertex (CSV) tagging algorithm~\cite{CMS-PAS-BTV-13-001,Chatrchyan:2012jua}.
Different working points are chosen for the various analyses: a loose working point providing an efficiency for \cPqb{}~quark jets of about 85\% and a light-flavor jet misidentification (mistag) rate of 10\%, a medium working point with 70\% \cPqb-quark jet efficiency and 1\% light-flavor jet mistag rate, and a tight working point with 50\% \cPqb-quark jet efficiency and 0.1\% light-flavor jet mistag rate.
Only jets with $\abs{\eta} < 2.4$ (within the CMS tracker acceptance) are identified  with this technique.

A number of Monte Carlo (MC) event generators are used to simulate the signal and backgrounds.
Signal events are produced with \MADGRAPH{} (v5.1.3.30)~\cite{madgraph2}, with a non-SM Yukawa coupling of $\Ct= -1$, and then passed through \PYTHIA{} (v6.426)~\cite{SimPythia} to add an underlying event and to perform parton showering and hadronization.
The masses for the top quark and Higgs boson are set to 173 and 125\GeV, respectively.
The CTEQ6L1~\cite{CTEQ} parton distribution function (PDF) set is used.
The sample is produced either using the five-flavor scheme or the four-flavor scheme.
Processes such as \ttbar\ plus additional particles (heavy-flavor jets, light-flavor jets, gluons, or bosons), \PW/\Z\ plus jets, and di- and tri-boson production are all generated with \MADGRAPH.
Single top quark plus jets and inclusive Higgs boson production are generated with \POWHEG{} (v1.0, r1380)~\cite{Frixione:2007vw,Alioli:2010xd}.
Both multijet (QCD) and \ttH{} production are simulated with \PYTHIA.\@
The detector response is simulated using a detailed description of the CMS detector based on the \GEANTfour{} package~\cite{GEANT4}.
All processes have been normalized to the most recent theoretical cross section computations.

The simulated samples are reweighted to represent the pileup distribution as measured in the data.
To match the performance of reconstructed objects between data and simulation, the latter is corrected with a set of data/MC scale factors.
Leptons are corrected for the difference in trigger efficiency, as well as in lepton identification and isolation efficiency.
Corrections accounting for residual differences between data and simulation are applied to the ECAL energy before combining the energy with the momentum determined from the tracker for electrons.
Similar corrections are applied to the muon momentum.

\section{Description of the analyses}
\label{sec:an_des}

The $t$-channel single top quark plus Higgs boson process has, at tree level, three particles in the final state: a top quark, a Higgs boson, and an additional quark jet, which tends to be emitted in the forward region.
A spectator b quark is produced through splitting of a gluon in the incoming proton, resulting in an additional bottom-flavor jet (b jet) that can enter the detector acceptance. Other Higgs boson production mechanisms, such as
\tth, are considered as background.
All of the analyses make use of the leptonic decay of the top quark, which yields a high-momentum lepton and an identifiable b jet.
Requiring these objects in the event improves the signal-to-background ratio for each analysis.
The analyses are distinguished by the Higgs boson decay channel, as
described in the following subsections.

\subsection{\texorpdfstring{$\PH\to\Pgg\Pgg$}{H to gamma gamma} channel}
\label{sec:dipho}

The diphoton branching fraction of the Higgs boson in the standard model is very small (0.23\%) but the diphoton final state allows very good background rejection thanks to the excellent diphoton invariant mass resolution provided by the CMS detector.
A negative top quark Yukawa coupling would not only enhance the yield of \tHq events, but also more than double the rate of Higgs bosons decaying to diphotons.
Thus the diphoton final state of the Higgs boson decay in \tHq events is
expected to be highly sensitive to the top quark Yukawa coupling.

The data for the diphoton analysis are collected using diphoton triggers with two different photon identification schemes. One requires calorimetric identification based on the electromagnetic shower shape and isolation of the photon candidate. The other requires only that the photon has a high value of the R9 shower shape variable, which is defined as the ratio of the energy contained in a 3$\times$3 array of ECAL crystals centered on the most energetic deposit in the supercluster to the energy of the whole supercluster.
The \et thresholds at trigger level are 26 (18)\GeV and 36 (22)\GeV on the leading (subleading) photon depending on the running period.
To maintain a high signal efficiency, trigger paths based on both photon identification schemes are combined in the offline data selection.

The event selection requires the presence of two photons, with the transverse momentum of the leading photon (${\pt}_1$) greater than $50\,\mgg/120$, where $\mgg$ is the reconstructed invariant mass of the diphoton system, and that of the subleading photon greater than 25\GeV.
The stringent requirement on ${\pt}_1$ is found to have very high efficiency~($>$98\%) for the signal and reduces the contributions of nonresonant backgrounds.
The presence of exactly one isolated electron or muon with $\pt > 10\GeV$ and at least one b quark jet with $\pt > 20\GeV$ are required to identify the leptonic decay of the top quark.
If more than one jet is \cPqb\ tagged, the one with the largest transverse
momentum is chosen as the \cPqb\ jet candidate from the top quark decay. Finally, the highest $\pt$ jet in the event that is not \cPqb\ tagged
must have $\pt > 20\GeV$ and $\abs{\eta} > 1$.

After applying these requirements, a multivariate method is used to further reduce the \ttH{} contribution.
A Bayes classifier, L, is constructed as the ratio of signal over signal plus background likelihoods for a chosen set of
discriminating observables:
\begin{equation}\label{eq:like1}
\mathrm{L}(\mathrm{x})=\frac{\mathrm{L}_\mathrm{S}(\mathrm{x})}{\mathrm{L}_\mathrm{S}(\mathrm{x})+\mathrm{L}_\mathrm{B}(\mathrm{x})}
\end{equation}
For each event the signal (L$_\mathrm{S}$) and background (L$_\mathrm{B}$) likelihoods are calculated as the product of the respective signal and background probability density functions ($p$),
evaluated at the observed values (x$^j$):
\begin{equation}\label{eq:like2}
 \mathrm{L}^i(\mathrm{x})=\prod_{j} p^i_{j}(\mathrm{x}^j),
\end{equation}
where $i$ stands for each signal or background process and $j$ for each variable considered.
The classifier is built from the following variables: the jet multiplicity
in the event; the transverse mass of the top quark using the lepton, the
candidate \cPqb\ jet and the missing transverse momenta;
the pseudorapidity of the light quark candidate; the rapidity gap between the lepton and the forward jet;
and the charge of the lepton candidate. The last observable is chosen as
the pp initial state is more likely to produce a top quark rather than an top antiquark.
All these variables are observed to discriminate well between simulated \tth{} and \thq{} events\,\cite{ref:Micheli}. The linear correlation coefficients for the input variables are all less than 10\% for both signal and background processes. The classifier value is required to be greater than 0.25, to suppress the \tth{} contribution to the signal sample.
This requirement retains about 90\% of the signal events.

The invariant mass of the diphoton system is the primary search variable
for a signal-like excess, as the signal would appear as a narrow diphoton
resonance centered at the known Higgs boson mass \mh = 125\GeV.

The backgrounds can be classified according to their resonant or nonresonant behavior in the diphoton system;
a different approach has been adopted to estimate the rate from each category. Resonant backgrounds
give rise to a Higgs boson decaying to two photons in the final state. These backgrounds are dominated by
the \tth process and also include Higgs production in association with a
vector boson (VH); they appear as an additional contribution under the expected signal peak, and are evaluated
using MC simulation. Nonresonant backgrounds are evaluated
from the \mgg{} sidebands. The main nonresonant background processes include diphoton production in association with jets~($\gamma\gamma$+jets),
single-photon production in association with jets~($\gamma$+jets), and
diphoton events produced in association with top quarks~(\ttgg, \tgg).

The signal region is defined as the
$\pm$3\GeV range around the nominal Higgs boson mass.
While the contribution of resonant backgrounds is taken from the simulation,
nonresonant backgrounds are evaluated by counting the events in the \mgg{}
sidebands $100\GeV < \mgg <  (m_{\PH}-3\GeV)$ and
$(m_{\PH}+3\GeV) < \mgg < 180\GeV$, which have negligible signal contamination.

\begin{figure}[th]
\centering
\includegraphics[width=0.4\textwidth]{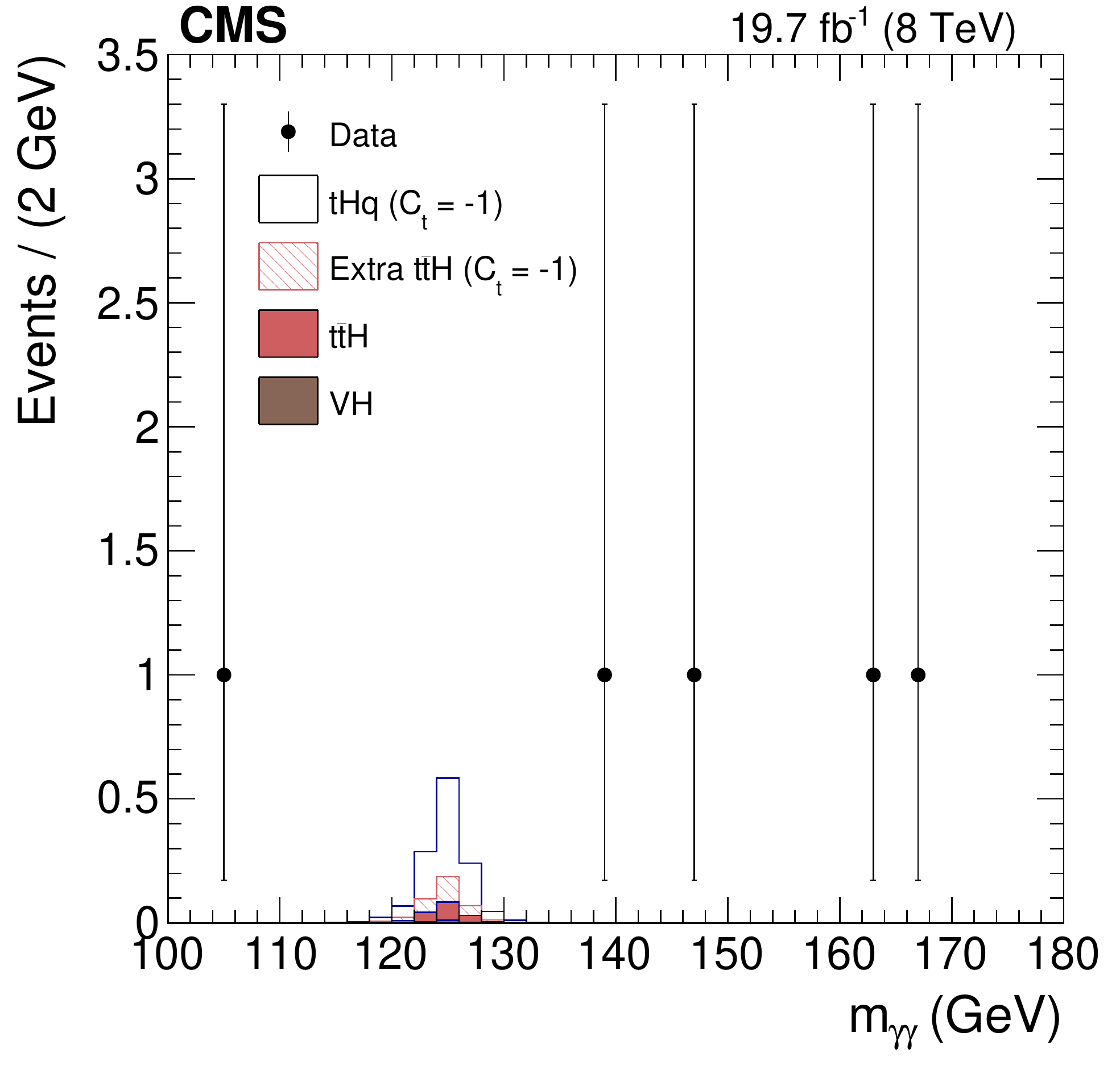}
\includegraphics[width=0.4\textwidth]{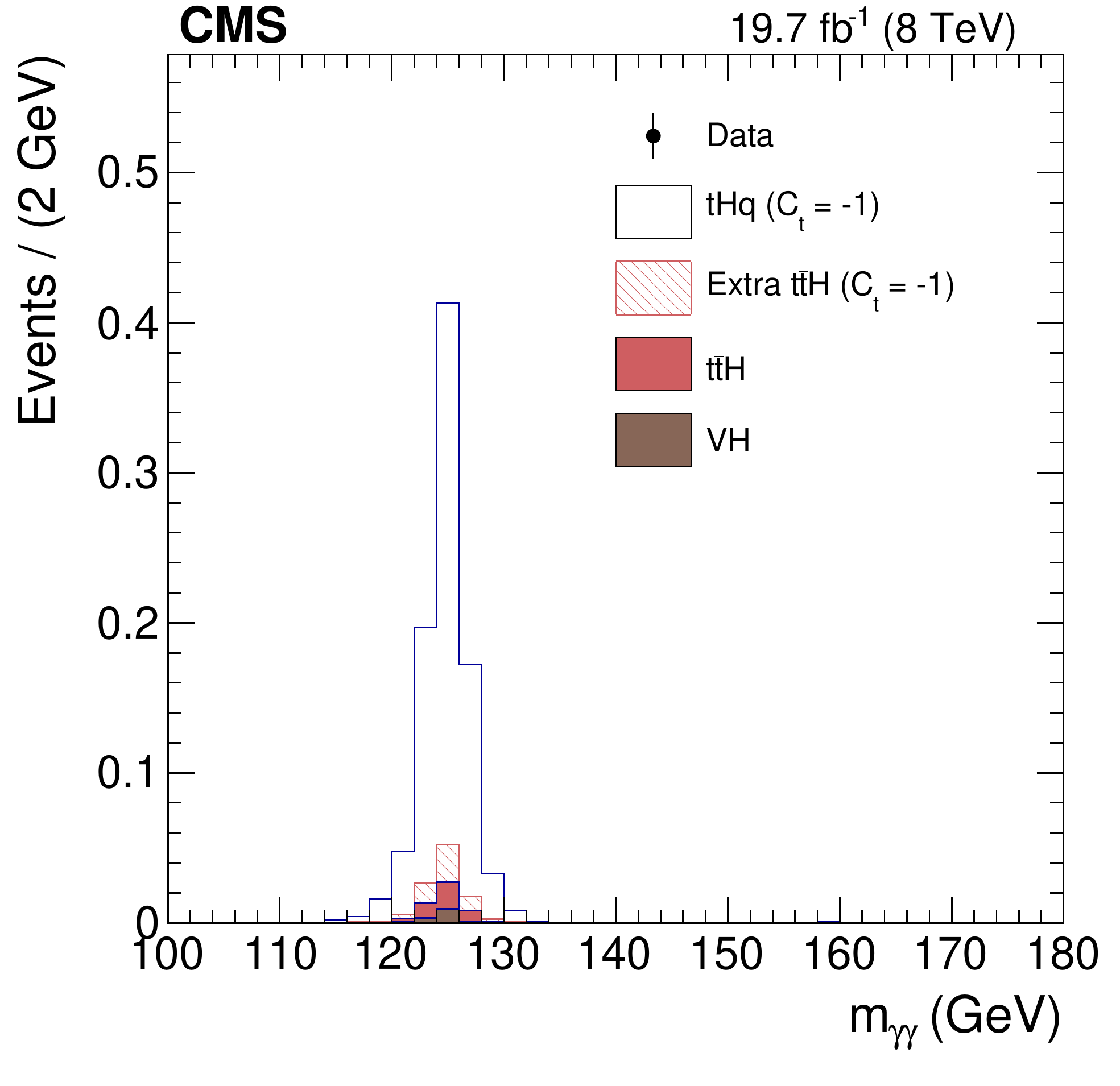}
\caption{Invariant mass of the diphoton system for events passing the event selection requirements, but for the likelihood discriminant cut (left), and for events passing the full selection (right).
The data~(black markers) are compared to the MC simulation~(stacked histograms).
No events are observed after the requirement on the likelihood discriminant.
\label{fig:mgg}}
\end{figure}

The event yields in the signal region are shown in
Table~\ref{tab:yields}. The selection has an expected efficiency of 17\% for \tHq{} events in the diphoton decay channel.
Figure~\ref{fig:mgg} shows the \mgg\ spectrum for events passing
the event selection before and after the likelihood requirement.

\begin{table}[tb]
\topcaption{Expected yields for the diphoton analysis, based on simulations.  Yields are counted for events with diphoton mass in the 122--128\GeV range.
The additional contributions to the \ttH\ and $\mathrm{VH}$  processes arising from the enhanced Higgs to diphoton branching fraction due to the $\Ct=-1$ assumption are marked with a dagger~($^{\dagger}$).}
\centering
\begin{tabular}{ll}\hline
          Process & Events                    \\ \hline
\tHq\ ($\Ct=-1$)     & 0.67                     \\ \hline
\tth{}            &  $0.03 + 0.05^{\dagger}$ \\
VH     &  $0.01 + 0.01^{\dagger}$ \\
Other \PH& 0                        \\
\hline
Data              & 0 \\
\hline
\end{tabular}
\label{tab:yields}
\end{table}

No events pass the selection. In order to model the nonresonant background shape using data,
a control region with relaxed \cPqb\ tagging requirements is defined.
The functional form chosen for the \mgg\ distribution of background
events is an exponential, and the uncertainty in the knowledge of the
background shape is assessed by defining an orthogonal control region in
which the isolation requirements on one of the two photons are inverted.
This uncertainty amounts to 33\%.  The number of events observed
and the systematic uncertainties are later used to set a limit on the rate of
\tHq production.
\subsection{\texorpdfstring{$\PH\to\cPqb\cPaqb$}{H to bbbar} channel}
\label{sec:bb}

The search for \tHq\ in the $\PH\to\cPqb\cPaqb$ decay final state benefits from the large Higgs to bottom-antibottom quarks branching fraction, but suffers from significant backgrounds from $\ttbar$ events.

\subsubsection{Event selection}\label{sec:selection_bbar}

The analysis is performed with data collected with two triggers:  one requiring an electron candidate with
$\pt > 27\GeV$ and $\abs{\eta} < 2.4$, the other requiring a muon candidate with $\pt > 24\GeV$ and $\abs{\eta} < 2.1$. In each case the lepton must be isolated.  The effect of the triggers is emulated in all simulated data sets. An event in the electron (muon) channel is required
to contain exactly one electron (muon) candidate with $\pt > 30~(26)\GeV$ and pass a set of identification criteria labeled as ``tight''. In order to reject Drell--Yan (DY) and other processes with multiple
prompt leptons, events are rejected if additional leptons exist that pass a looser criterion.

The signal final state in this channel is expected to contain at least five quarks: two \cPqb\ quarks from the Higgs boson decay, one \cPqb\ quark each from the top quark decay and from the strong interaction, and a forward light quark from the $t$-channel process. Each event is thus required to contain at least four jets with $\pt > 30\GeV$
and the threshold for counting additional jets beyond the fourth is chosen to be 20\GeV. Jets with $\abs{\eta} > 2.4$ are considered only if they have $\pt > 40\GeV$.  A tight working point of the CSV \cPqb\ tagging algorithm is chosen to suppress the large background from top quark pair production, which contains a smaller number of genuine \cPqb\ quarks than the signal process.
This working point has typical tagging efficiencies of 55\% for \cPqb\ jets and 0.1\% for light-flavor jets.  To reject multijet events, a missing transverse energy selection is applied with thresholds optimized per channel:  $\MET > 45\GeV$ in the electron channel and $\MET > 35\GeV$ in the muon channel.  As the \cPqb\ quark
produced in the strong interaction of the \tHq process is often forward and falls outside the acceptance of the detector, two analysis samples are defined: one of events containing at least four jets with three of them
\cPqb-tagged and one of events containing at least five jets with four of them \cPqb-tagged.  Additionally, a two-tag control sample dominated by \ttbar\ plus jets events is used for validation of event reconstruction and signal extraction techniques described in the following section.

After this event selection is applied, the sample is dominated by the \ttbar\ plus jets background as well as other background contributions\,\cite{ref:Popov}.  The three-tag sample has an expected signal-to-background ratio of 0.7\%.  The four-tag sample has an improved ratio of approximately 2\% but suffers from a limited number of events. The background kinematic distributions and normalizations are taken from simulation and are adjusted in the final fit, taking into account all systematic uncertainties, which are described in more detail in Section~\ref{subsec:common_systs}. A cross-check approach that uses control data samples to model the dominant \ttbar\ plus jets background in the signal regions by employing \cPqb-tagging and mistagging efficiencies in the two-tag control sample gives consistent results.

\subsubsection{Event reconstruction under \texorpdfstring{\tHq\ and \ttbar}{tHq and ttbar} hypotheses}\label{sec:reconstruction_bbar}

The selected samples are dominated by \ttbar\ plus jets production, as shown in Sec.~\ref{sec:classification_bbar}. An artificial neural network (NN) is employed to separate the signal process from background, based on the features of \tHq and \ttbar\ plus jets events.
Prior to this, a correspondence between reconstructed jets and the final-state objects must be built in order to define the input variables to the NN.\@ For this purpose, each event is reconstructed under two hypotheses: (1) that it is a \tHq\ signal event, or (2) that it is a \ttbar\ plus jets background event.
Simulated events are used to assess the correctness of the assignment of jets to quarks.

For the jet assignment under the \tHq hypothesis in a simulated \tHq event, all possible ways to assign four reconstructed jets to the four final state quarks from $\PQt\PH\PQq\to3\PQb\PQq\ell\nu$ are considered,
where a correct event interpretation is present in the case where four jets can be matched to the appropriate quarks within a cone of radius $\Delta R=\sqrt{\smash[b]{(\Delta \eta)^2 + (\Delta \phi)^2}}=0.3$. If the distance
between at least one quark and its assigned jet is larger than this threshold, the event interpretation is flagged as wrong. The total number of possible interpretations is reduced by additional requirements:
because of \cPqb\ tagging considerations, \cPqb\ quarks can only be associated with central jets ($\abs{\eta}<2.4$), while only a jet failing the \cPqb\ tagging requirement can be assigned to the light recoil quark.

A NN is trained on \tHq\ events to distinguish between correct and wrong interpretations with variables employing kinematic characteristics of the signal, like the \pt of the softest jet from the Higgs boson decay, the $\abs{\eta}$ of the recoil jet, and the $\Delta R$ between the reconstructed top quark and the Higgs boson. Other variables include information such as \cPqb~tagging or the reconstructed jet charge.
The interpretation chosen for use in the analysis is the one that gives the largest NN response from all possible \tHq\ jet assignments.

Similarly, another NN is used for the interpretation of events under the assumption that they originate from semileptonic \ttbar decays. The NN is trained with $\ttbar\to2\PQb2\PQq\ell\nu$  simulated events, using both correct and wrong quark jet assignments in analogy with the \tHq jet assignment described above. The number of possible jet-quark combinations is restricted by requiring that only \cPqb-tagged jets can be assigned to the two \cPqb\ quarks.
The set of variables used under a \ttbar event interpretation is similar to the one of the tHq event interpretation. It makes use of kinematic relations between objects, such as the $\Delta R$ between the \cPqb\ and \PW{} boson from the hadronically decaying top quark, or the difference between the reconstructed top quark mass and \PW{} boson mass in the hadronic top quark decay. It also employs \cPqb\ tagging information and relations between the jet and lepton charges.
The jet assignment yielding the largest NN response is chosen as the event interpretation under the \ttbar hypothesis. Additional details regarding the event interpretation can be found in Ref.\,\cite{ref:Popov}.

\subsubsection{Event classification and signal extraction}\label{sec:classification_bbar}

The \tHq\ and \ttbar\ plus jets reconstruction algorithms described above are carried out on every event passing the selection criteria.  This allows the construction of two sets of observables, where one set describes the event under the \tHq\ hypothesis and the other the event under the \ttbar hypothesis.

These two sets, together with the lepton charge, form the list of input variables for the final NN, which classifies events as signal- or background-like: $\abs{\eta} $ of the recoil jet; number of \cPqb-tagged jets among the two jets from the Higgs boson decay; \pt\ of the Higgs boson; \pt\ of the recoil jet; $\Delta R$ between the two light-flavor jets from the hadronic top quark decay; reconstructed mass of the hadronically decaying top quark; number of \cPqb-tagged jets among the two light-flavor jets from the hadronic decay of the top quark; and lepton charge\,\cite{ref:Popov}.

\begin{figure}[!htb]
  \centering
    \includegraphics[scale=0.39]{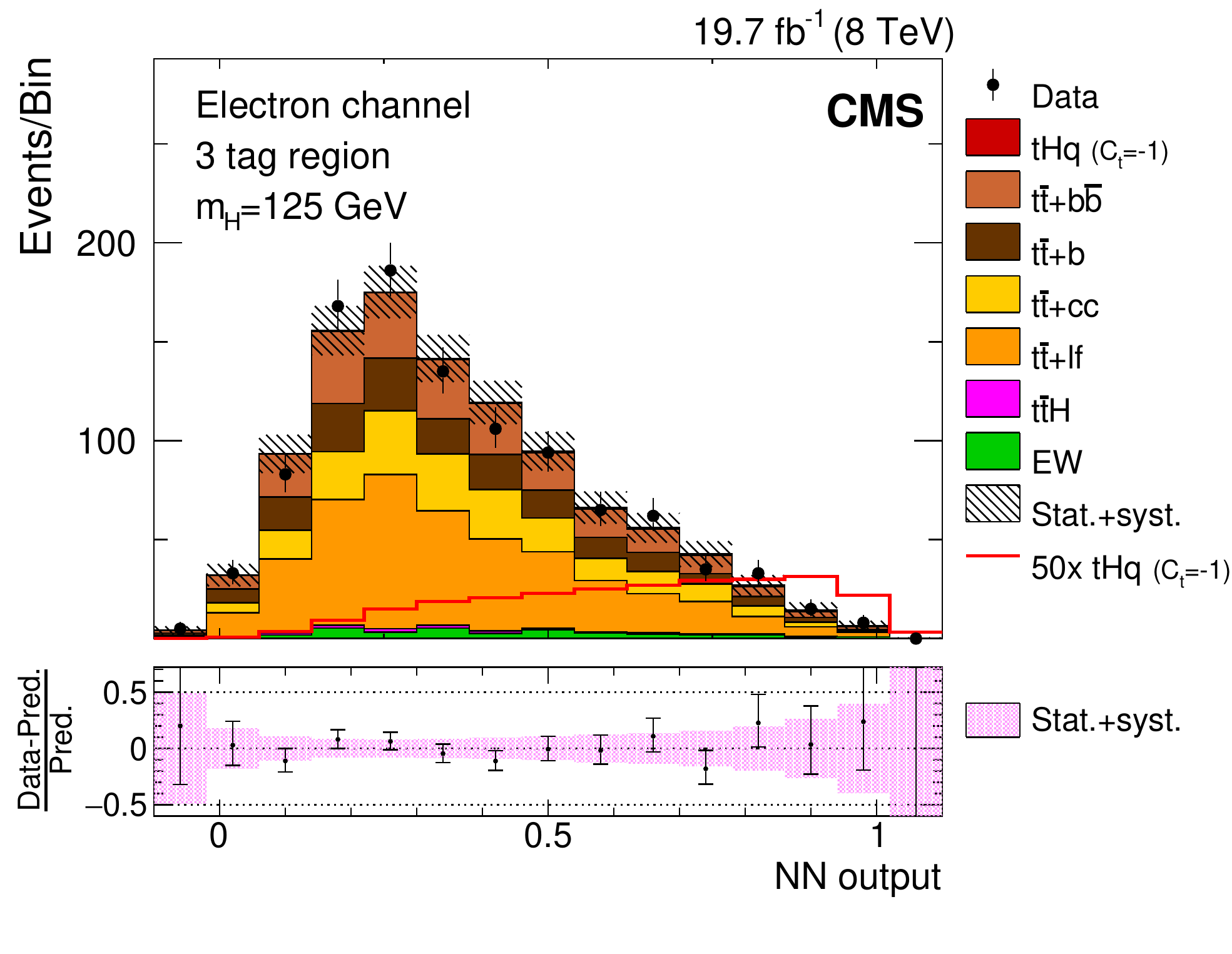}
    \includegraphics[scale=0.39]{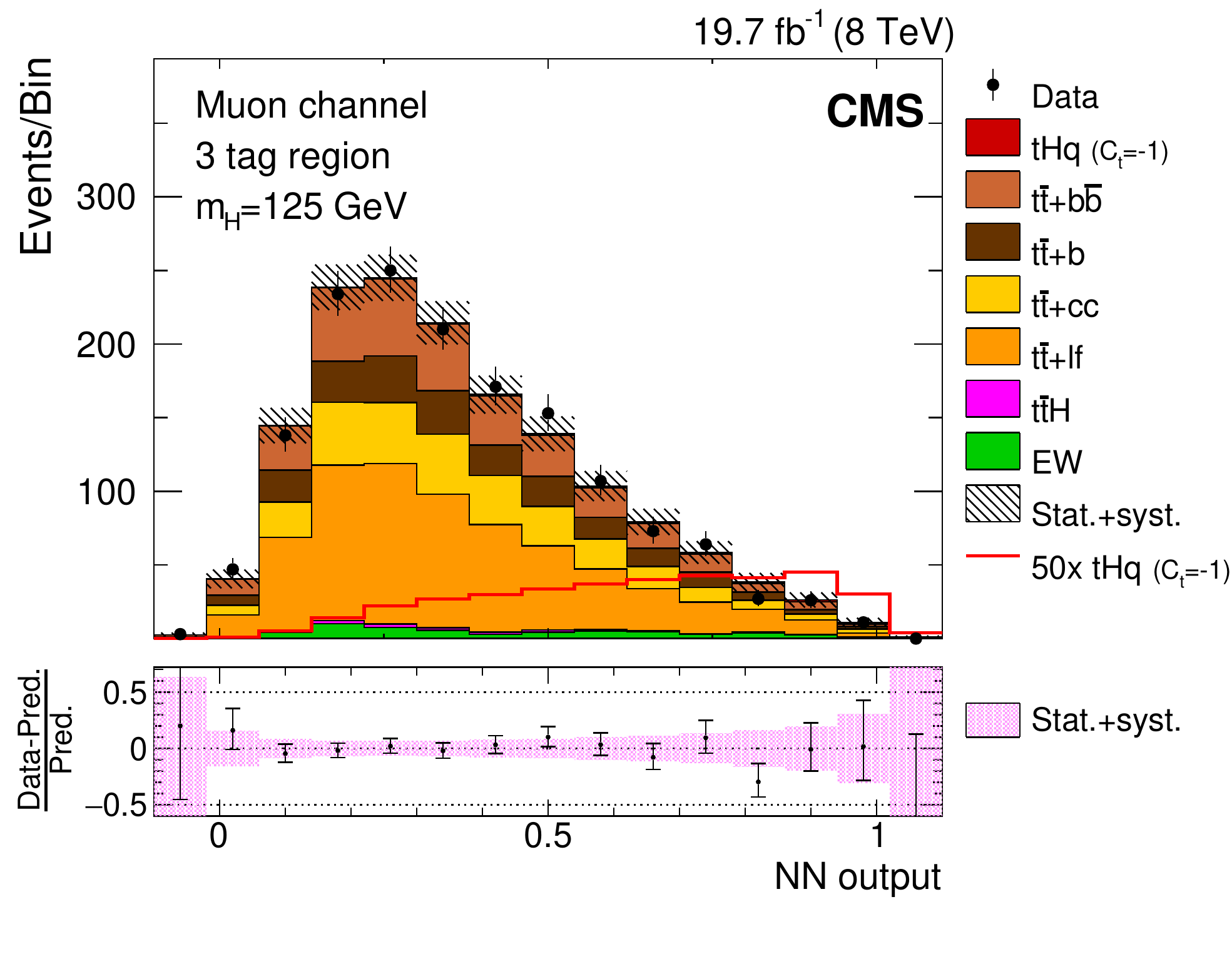}\\
    \includegraphics[scale=0.39]{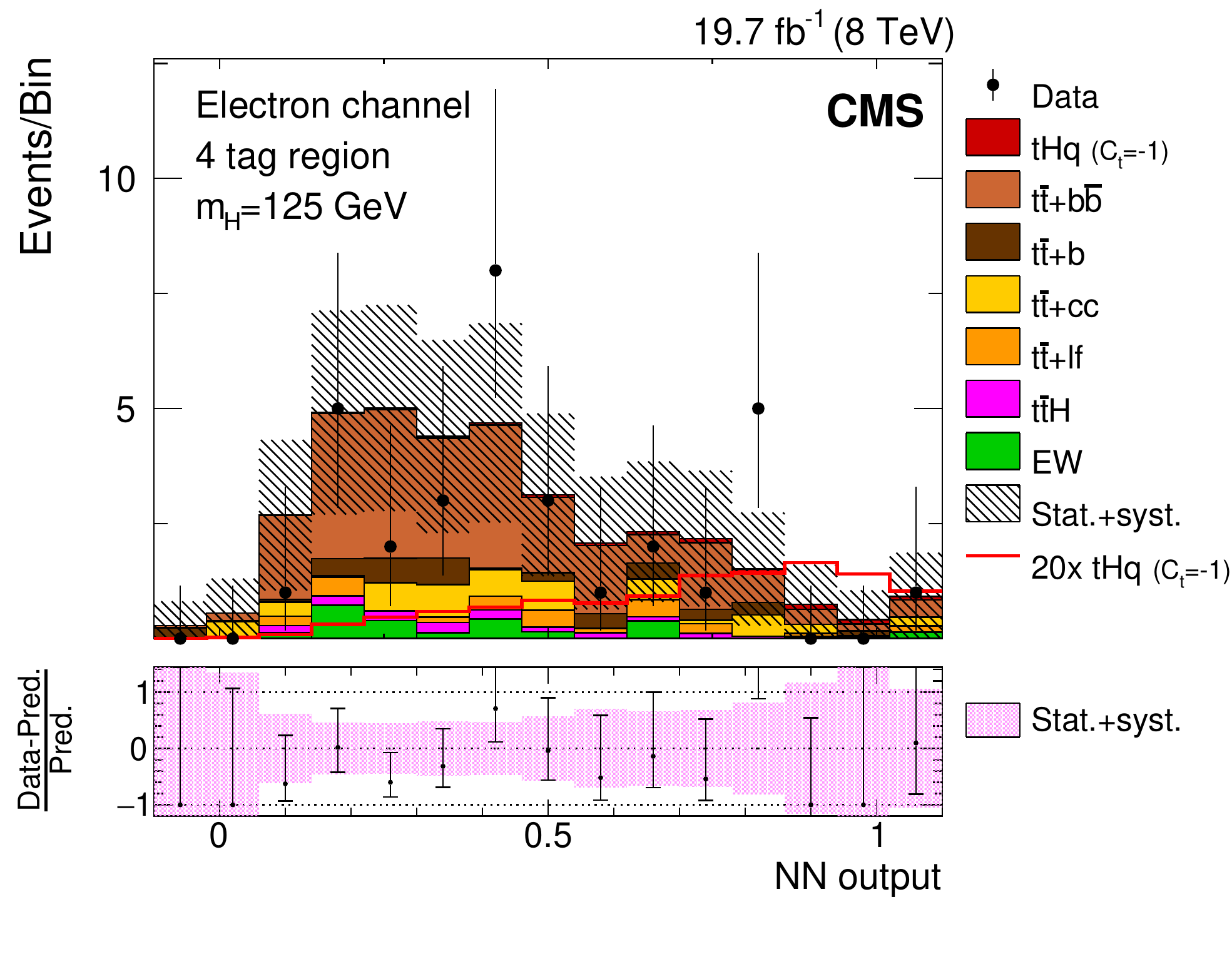}
    \includegraphics[scale=0.39]{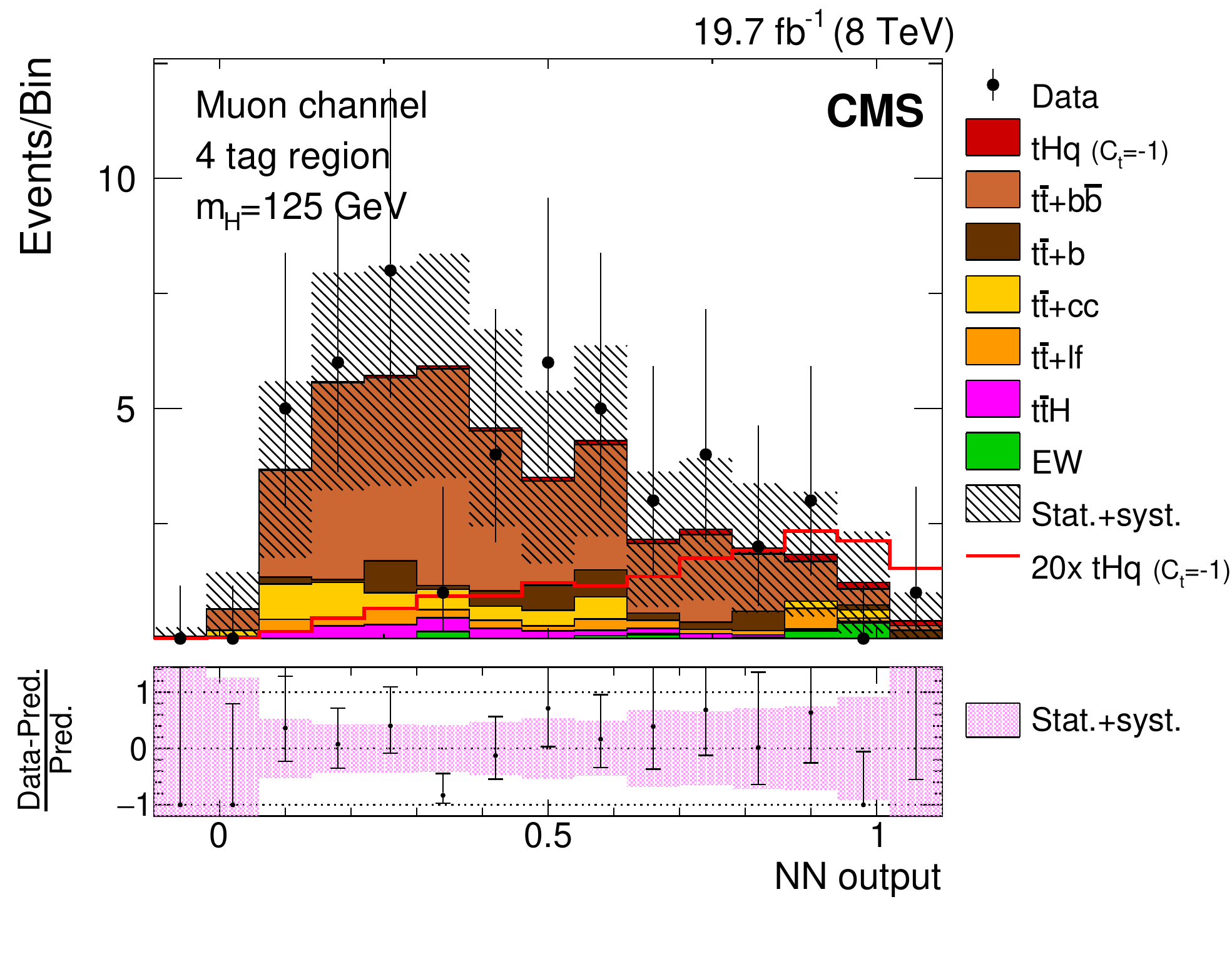}\\
\caption{Distributions of the NN output for the $\Hbb$ channel for events with three (four) b-tagged jets are shown in the upper (lower) row. The left (right) column shows events containing a high-$\pt$ electron (muon). All backgrounds are normalized to the output of a maximum likelihood fit of the corresponding distributions. ``EW'' indicates electroweak backgrounds: single top quark, \PW/\Z\ boson plus jets, and di- and tri-boson production.  The line shows the expected contribution from the \tHq\ process with $\Ct = -1$ multiplied by the factor indicated in the legend.  In the box below each distribution, the ratio of the observed and predicted event yields is shown. The shaded band represents the post-fit systematic and statistical uncertainties.
\label{fig:mva_postfit_bbar}}
\end{figure}

Figure~\ref{fig:mva_postfit_bbar} shows the distributions of the final event classifier in the three-tag and four-tag samples, separated by lepton flavor. The distributions of the NN outputs are used to extract the signal and to derive the upper limit on the cross section for \tHq production.
The normalizations of the distributions are taken from the result of a maximum likelihood fit where each background and the signal process are allowed to float within the assigned systematic and statistical uncertainties. The resulting distributions show a good agreement with data and residual differences are well covered by the total uncertainties.

\subsection{\texorpdfstring{\HWW}{HWW} channel}
\label{sec:leps}

The Higgs boson decay to two W bosons (with one boson off-shell) has the second-largest branching fraction in the standard model. The associated tHq, \HWW\ and $\PQt\to\PW\cPqb$ final state allows several combinations of leptonically and/or hadronically decaying W bosons.
Two channels are exploited here, in which either all three \PW\ bosons decay leptonically, or the pair of \PW\ bosons with equal charge, resulting in a signature of either three leptons (electrons or muons), or two same-sign leptons with two light quark jets. The \tHW process can also result in this set of leptons.  In the \tHq process both the tri- and dilepton signatures are accompanied by a \cPqb\ quark and a light-flavor forward jet.  In addition, a significant \MET can be expected because of the undetected neutrinos from the leptonic W decays.  While the leptonic branching fraction of the W is relatively small, the presence of multiple leptons and identified b jets in the final state reduces the number of background events.  The \tHq search in this final state has some acceptance for events where \Pgt\ leptons, stemming either from the decay of one or more W bosons, or from Higgs boson decays, give rise to electrons or muons in the decay chain. Events with hadronically decaying \Pgt\ leptons are considered separately in Section~\ref{sec:taus}.

The trigger used to select the analysis sample requires the presence of two high-\pt\ electrons or muons. The \pt
thresholds are 17 and 8\GeV for the leading and subleading leptons, respectively. The trigger efficiency for signal events with two high-\pt leptons is higher than 98\%, and almost 100\% for those with three leptons.

Various SM processes contribute as background in the signal region: diboson (\WZ, \WW, \ZZ, \WWqq) and triboson (WWW, WWZ, WZZ) production, associated
production of \ttbar with a boson (\ttW, \ttZ, \ttH, \ttG, and \ttGStar), \ttbar with two \W bosons (\ttWW), single top quark associated production with a \Z boson
(\tZq), and production of same-sign \W bosons via double parton scattering (\WW).

\subsubsection{Background modeling}\label{sec:background_multileptons}
While diboson backgrounds are produced with a relatively large cross section, their contribution is strongly reduced by imposing a veto on lepton pairs compatible with a \Z boson decay in the trilepton channel (\Z\ boson veto) or by vetoing additional leptons in the event in the same-sign dilepton channel. Furthermore, the requirement of a \cPqb-tagged or forward jet suppresses contributions from diboson
processes. The rate of DY events is strongly reduced by the \Z\ boson veto (trilepton channel) and the third lepton veto (same-sign dilepton channel).
Triboson production has a very small cross section and is further reduced by rejecting events with extra leptons. In addition, both diboson and triboson
production do not generally include forward jets or jets from \cPqb{}~quark decays.

Associated production of \ttbar and vector bosons (\ttW, \ttZ) or Higgs boson (\ttH), although having fairly small cross sections, have high lepton and jet multiplicities as well as two final-state \cPqb\ quarks and can contribute significantly.
Because of its very large cross section, \ttbar production is expected to be the major source of background for both channels, when additional leptons are produced in the decay of \PB{} hadrons or when light jets are misidentified as leptons.

An additional background in the case of same-sign dileptons arises when the charge of a lepton in events with an opposite-sign lepton pair is misidentified.
This happens not so much because of track misreconstruction, but rather because of strongly asymmetric conversions of hard bremsstrahlung photons emitted from the initial lepton, and is therefore much more likely to occur for electrons than for muons. In the case where the original electron loses most of its energy to the radiated
photon, and the conversion daughter with opposite charge carries most of the momentum, the resulting track can have opposite curvature to the original
lepton. Furthermore, the same-sign channel has a contribution from the associated production of two same-sign \W bosons and two light quark jets, \WWqq.

Backgrounds involving nonprompt leptons and charge misidentification are estimated using data-driven methods. All the remaining processes are
estimated from MC simulation, corrected for data/MC scale factors and pileup distribution, using NLO cross sections where available.

\sloppy{
The ``tight-to-loose'' method is used for estimating the \ttbar background. It is based on defining two lepton selection levels: the tight criteria, corresponding to the full lepton identification used in the signal selection; and a looser selection designed to accept more background leptons.
The probability of a nonprompt lepton to pass the tight cut after passing the loose cut (the tight-to-loose rate, $f$) is then extracted from data control samples.
Nonprompt leptons include real leptons from heavy flavor hadron decays, jets from light quarks misreconstructed as leptons, as well as photon conversions.
Finally, the signal selection is extended with the loose lepton selection, and the additional event yield is weighted to arrive at an estimate for the expected contribution from nonprompt leptons.
Events with one tight and one loose lepton obtain a weight of $f/(1-f)$, whereas events with two loose leptons are weighted by \mbox{$- f_1 \, f_2 / \big((1-f_1)(1-f_2)\big)$}.
The method assumes $f$ to be consistent between signal and control samples, and that there are only two categories of leptons with consistent efficiencies of passing the tight selection: prompt leptons from \W and \Z boson decays, and nonprompt leptons.
}

The tight-to-loose rate is defined as the ratio between $N_\text{tight}$ and $N_\text{loose}$, where $N_\text{loose}$ is the number of candidate leptons that pass the loose selection, based on relaxed isolation and impact parameter requirements, and $N_\text{tight}$ is the number of loose leptons that also fulfill the tight requirements defined in the analysis. The rate $f$ is measured in a data sample enriched in background leptons, and parametrized as a function of the \pt, $\eta$, and lepton flavor.
For the lepton selections used, the electron tight-to-loose rate varies in the range 1--13\%, whereas the muon rate varies
between 5--23\%.

Similarly, the contribution of events with a misidentified lepton charge to the same-sign channel is estimated using the charge misidentification probability and the yield of opposite-sign pairs in the signal selection.
The electron charge misidentification probability is extracted from an independent data sample based on \Z boson decays, and cross checked with expectations from MC simulation. It is binned in \pt and $\eta$, and ranges from about 0.03\% in the barrel to between 0.08\% and
0.28\% in the endcap.
The muon charge misidentification probability in the relevant \pt range is  negligible.

\subsubsection{Event selection and signal extraction}\label{sec:selection_multileptons}
A relatively loose selection is applied to maintain a large signal efficiency while suppressing the main backgrounds. For the dilepton analysis, the presence of two same-sign leptons with $\pt > 20\GeV$ and invariant mass $m_{\ell\ell} > 20\GeV$ is required. No additional leptons can be present in the event. At least one
central jet with $\pt > 25\GeV$ is required to be tagged with the CSV algorithm using a loose working point. The event must also contain at least one forward jet ($\abs{\eta} > 1.0$)
and an additional central jet ($\abs{\eta} < 1.0$), both with $\pt > 25\GeV$.

For the trilepton analysis, the thresholds for the three lepton transverse momenta are $\pt > 20$, 10, and 10\GeV.
To suppress contamination from DY events, the reconstructed dilepton invariant mass closest to the \Z boson mass ($m_\Z$) must respect the constraint $\abs{m_{\ell\ell} - m_\Z} > 15\GeV$. The presence of large missing transverse energy suggests the presence of
multiple neutrinos, hence a cut on $\MET > 30\GeV$ is applied. Only events with one jet with $\pt > 25\GeV$ and $\abs{\eta} < 2.4$ tagged with the medium working point of
the CSV algorithm are selected, and at least one forward jet with $\pt > 25\GeV$ and $\abs{\eta} > 1.5$ must be present.

 The production cross section times branching fraction for the signal
(assuming $\Ct=-1$) is just a few fb, resulting in a fairly small signal-to-background fraction even for a tight selection.
Therefore a multivariate analysis method is used to build a Bayes classifier as in Eqs.~(\ref{eq:like1}) and~(\ref{eq:like2}) to further reduce backgrounds.
A search for an optimal set of variables is performed, to find those that best separate the signal and the backgrounds. The discriminating variables can be put into three broad categories: forward activity, jet and \cPqb\ jet multiplicity, and lepton kinematic properties and charge. The variables to enter the classifier are chosen to be minimally correlated while providing good discrimination power.

For the same-sign lepton final state, the following set of variables has been chosen: the scalar sum of the \pt of all the jets; the jet multiplicity; the medium \cPqb-tagged jet multiplicity; the $\abs{\eta}$ value of the leading jet with $\abs{\eta} > 1.0$;
the  $\Delta\eta$ value between the most forward jet and second-most forward jet or lepton; the charge of the leptons; the azimuthal angle difference between the two leptons ($\Delta\phi_{\ell\ell}$); and the \pt of the trailing lepton.

In the case of the trilepton final state the selected variables are: the multiplicity of untagged central jets (with $\abs{\eta} < 1.5$); the number of forward jets with $\abs{\eta} > 2.4$; the total sum of the charges of the three leptons; the minimum value of $\Delta R$ between the leptons in the event; and the $\Delta\eta$ value between the \cPqb-tagged jet and the most forward jet.

\begin{figure}[!htb]
  \centering
    \includegraphics[scale=0.26]{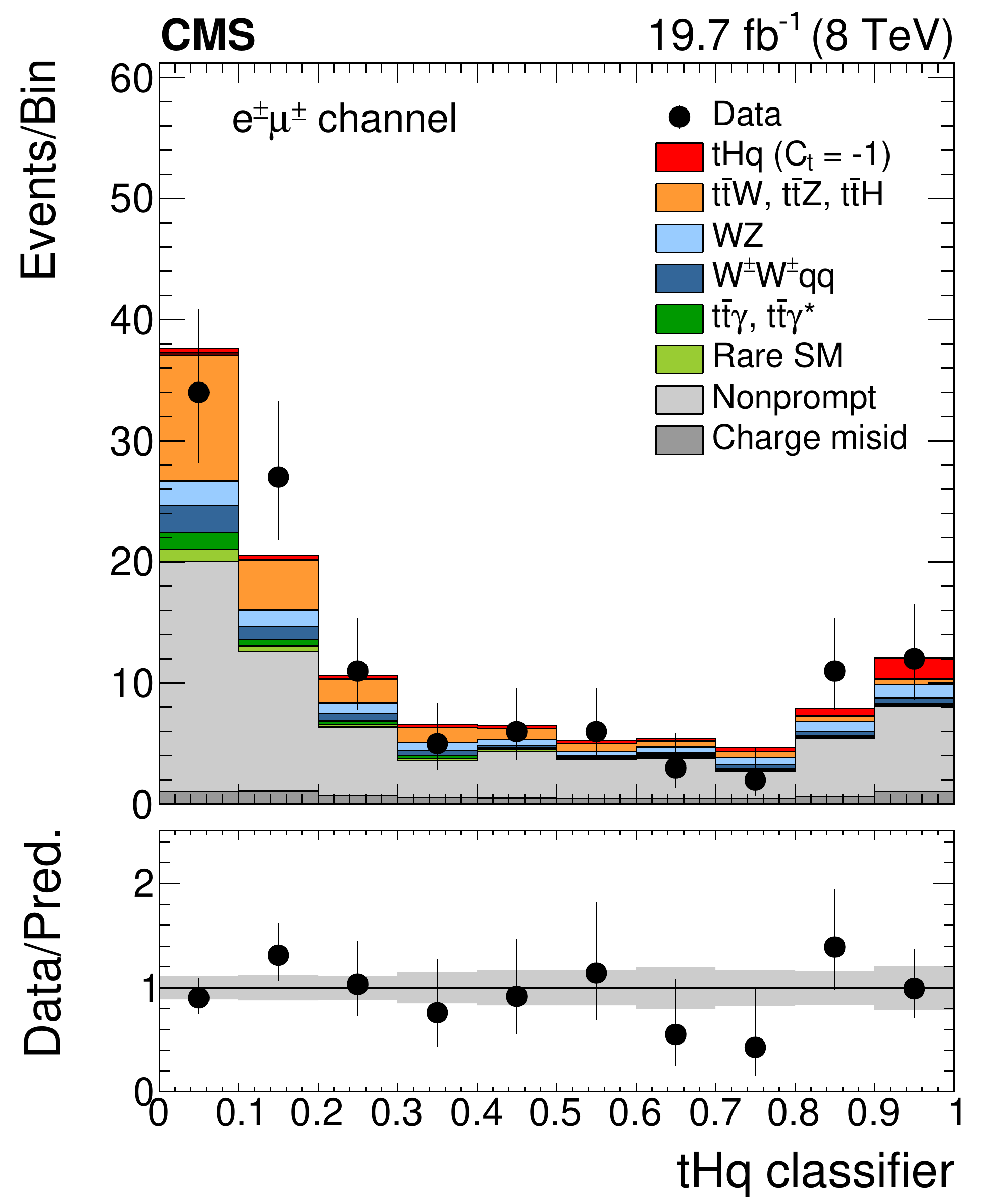}
    \includegraphics[scale=0.26]{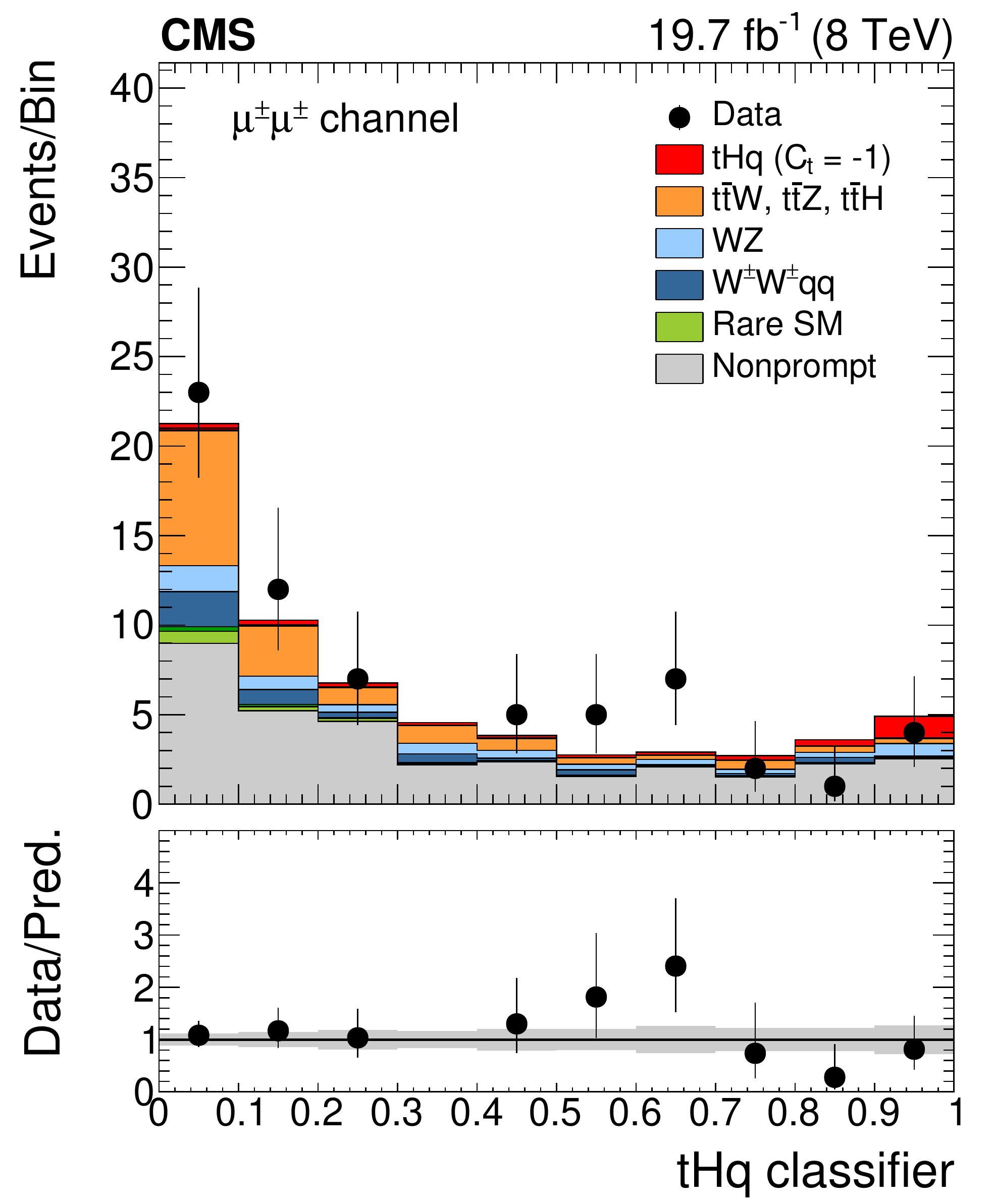}
    \includegraphics[scale=0.26]{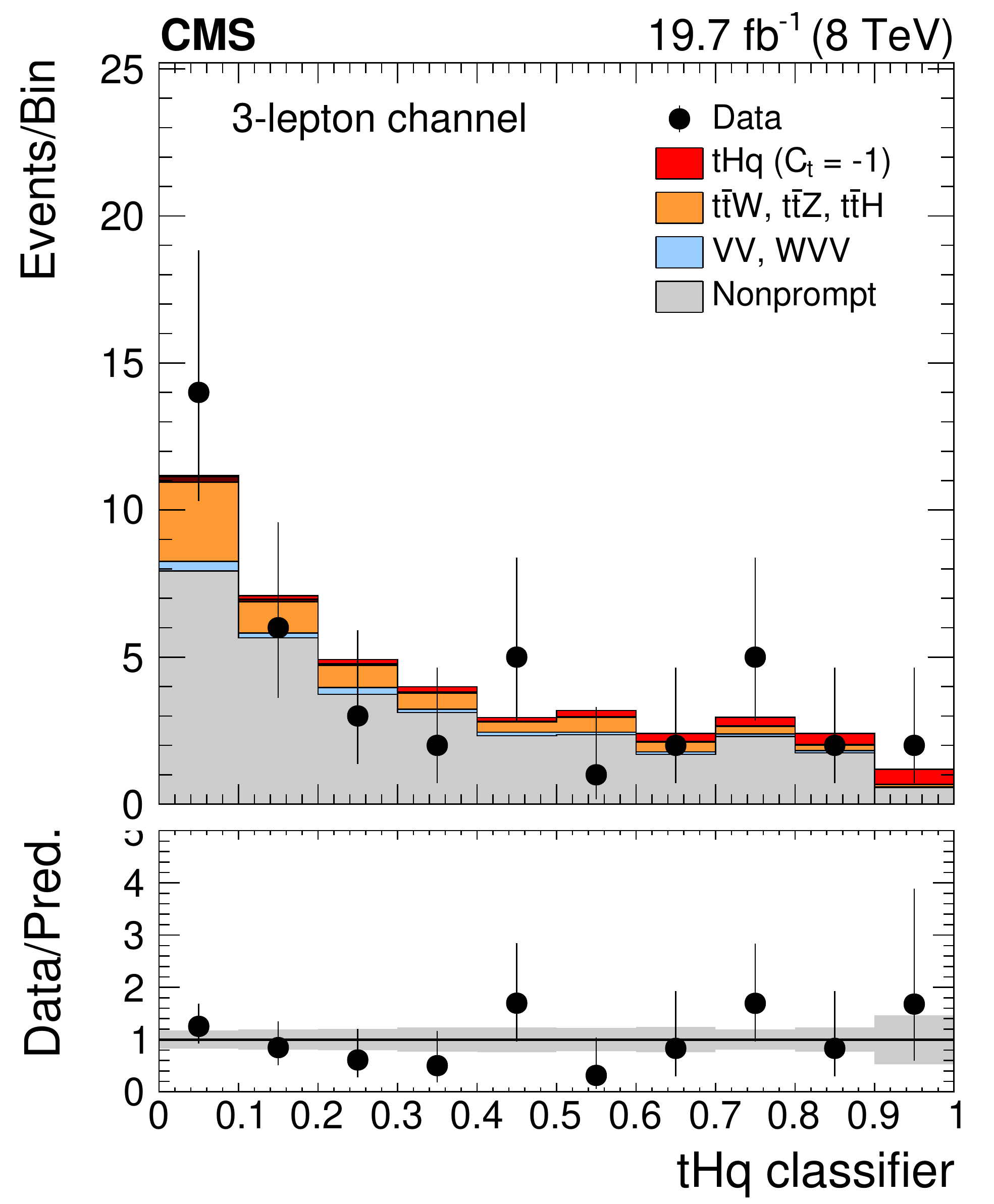}
  \caption{Post-fit Bayes classifier output, for the \emu (left), \mumu (center), and trilepton channel (right).
  In the box below each distribution, the ratio of the observed and predicted event yields is shown.
  The gray band represents the post-fit systematic and statistical uncertainties.\label{fig:postfitshapes}}
\end{figure}

To derive an upper limit on the signal production cross section for $\Ct=-1$, a maximum likelihood fit of the classifier output is then performed in all three channels.
Table~\ref{tab:postfityields} shows the observed data yields and the post-fit expected number of signal
and background events, where the trilepton channel, $\ell\ell\ell$, consists of eee, $\Pe\Pe\mu$, $\mu\mu\Pe$, and $\mu\mu\mu$ final states. The post-fit classifier output for all the channels is shown in  Fig.~\ref{fig:postfitshapes}.

\begin{table}[!htb]
\topcaption{Data yields and post-fit expected backgrounds after the event pre-selection for single top plus Higgs events appearing in events with \emu, \mumu, or $\ell\ell\ell$. Contributions from tHq and tHW are shown separately, as well as expected events where the Higgs boson decays to W bosons, or to tau leptons. Uncertainties include systematic and statistical sources. ``Rare SM'' comprises \VVV, \tbZ, \ZZ,  \ttWW, and \WW processes for the dilepton channels, and \WVV for the trilepton channel.}
  \centering
      \begin{tabular}{rccc}
        \hline
        Process          & \emu                 & \mumu               & $\ell\ell\ell$    \\
        \hline
        \tHWtt ($\Ct=-1$)           & $  0.13 \pm  0.14 $  & $  0.10 \pm  0.1\
2 $ & $0.12 \pm 0.12 $ \\
        \tHWWW ($\Ct=-1$)          & $  0.47 \pm  0.48 $  & $  0.28 \pm  0.29\
 $ & $0.35 \pm 0.35 $ \\
        \tHqtt ($\Ct=-1$)          & $  0.90 \pm  0.91 $  & $  0.59 \pm  0.61\
 $ & $0.56 \pm 0.58 $ \\
        \tHqWW ($\Ct=-1$)          & $  3.73 \pm  3.84 $  & $  2.55 \pm  2.62\
 $ & $1.73 \pm 1.80 $ \\
         \hline
        Total signal ($\Ct=-1$)    & $  5.22 \pm  3.98 $  & $ 3.53 \pm  2.71  $ & $2.76 \pm  1.93$ \\
      \hline
        \WWqq            & $  6.03 \pm  0.85 $  & $  4.60 \pm  0.68 $ & ---                \\
        \WZ, \WW, \ZZ    & $  8.83 \pm  3.25 $  & $  5.47 \pm  2.10 $ & $1.19 \pm 0.14$  \\
        Rare SM bkg.     & $  2.57 \pm  1.23 $  & $  1.40 \pm  0.68 $ & $0.11 \pm 0.03$  \\
        \ttGStar         & $  1.04 \pm  0.42 $  & $  0.50 \pm  0.20 $ & ---               \\
        \ttG             & $  2.02 \pm  0.60 $  & $  0.09 \pm  0.03 $ & ---               \\
        \ttZ             & $  2.87 \pm  0.50 $  & $  2.23 \pm  0.41 $ & $2.21 \pm 0.36 $ \\
        \ttW             & $ 14.85 \pm  3.32 $  & $ 10.18 \pm  2.24 $ & $3.03 \pm 0.51 $ \\
        \ttH             & $  3.24 \pm  0.47 $  & $  2.26 \pm  0.34 $ & $1.52 \pm 0.18 $ \\
        Charge misid     & $  6.96 \pm  1.76 $  & ---                  & ---                \\
        Nonprompt        & $ 63.7  \pm 12.5  $  & $ 33.3  \pm  8.3  $ & $31.4  \pm 6.5  $ \\
        \hline
        Total background & $ 112.1  \pm 13.5  $ & $ 60.1  \pm 9.0  $  & $ 39.5  \pm  6.6  $ \\
        \hline
        Data             & 117                  & 66                  & 42\\
        \hline
      \end{tabular}
    \label{tab:postfityields}
\end{table}
\subsection{\texorpdfstring{$\PH\to\Pgt^+\Pgt^-$}{H to tau+ tau-} channel}
\label{sec:taus}

The previous section presented a strategy for identifying \tHq events that captured events with leptonically decaying tau leptons.  An orthogonal strategy is devised to analyze signal events with a reconstructed tau lepton, through its decay to hadrons (\tauh).
The analysis described here is based on two final states with three reconstructed leptons, $\Pe\mu\tauh$ and $\mu\mu\tauh$.
These final states are chosen to select signal events where the two $\Pgt$ leptons from the Higgs boson decay give rise to an $\Pe\tauh$ or $\Pgm\tauh$ final state and the top quark decay produces the third lepton.

The major SM background processes that can lead to the same lepton final state include $\PW\Z$, $\Z\Z$, \ttH, and $\ttbar+\PW/\Z$ production.
The contributions from reducible backgrounds include $\ttbar$, single top, $\PW+$jets, $\Z$+jets, and multijet production.
These contributions are estimated using events from control samples in the data.

\subsubsection{Event selection}
\label{sec:selection_tautau}

Candidate events are collected using either $\Pe\Pgm$ or $\Pgm\Pgm$ triggers, depending on the final state.
In the final event selection, the leading (subleading) electron or muon is required to have $\pt^\ell > 20\,(10)\GeV$.
Electrons (muons) are required to have $\abs{\eta} < 2.5\,(2.4)$ and to pass basic identification requirements.
To suppress secondary leptons from b-flavored hadron decays, an isolation classifier is computed with boosted decision trees using variables based on impact parameters with respect to the reconstructed primary interaction vertex (defined as the vertex with highest $\sum{\pt^2}$ of its associated tracks), variables related to the isolation of the lepton, and variables related to the reconstructed jet closest to each lepton~\cite{CMS:tthcombo_paper}.

In addition, either the electron and muon in the $\Pe\Pgm\tauh$ final state or the two muons in the $\Pgm\Pgm\tauh$ final state are required to have equal charges.
This same-sign requirement suppresses contributions from backgrounds with prompt opposite-sign dileptons and additional jets that can be misidentified as hadronic tau leptons, such as $\Z/\Pgg^{*} \to \Pgmp \Pgmm + \text{fake \Pgt\ jet}$ and $\ttbar/\Z\to\Pgt\Pgt+  \text{fake \Pgt\ jet}$.

The selected leptons ($\Pe$, $\Pgm$, and $\tauh$) are required to be at least 0.5 apart in $\Delta R$.
To reduce contributions from \ZZ~and \ttbar\Z backgrounds, events with additional isolated electrons and muons are rejected.

The $\tauh$ candidate, reconstructed as described in Sec.\,2, is required to have $\pt > 20\GeV$, $\abs{\eta} < 2.3$, and to pass identification and isolation criteria to reject misidentified $\tauh$ candidates from jets, electrons, or muons~\cite{Chatrchyan:2014nva}.
The charge of the $\tauh$ candidate is required to be opposite to that of other leptons, \Pe\ or \Pgm.
To suppress background events without \cPqb~quark jets,
the presence of at least one jet with $\pt>20$\GeV and $\abs{\eta}<2.4$ identified as coming from a \cPqb\ quark with the medium working point of the CSV algorithm is required.
This requirement particularly reduces the contamination from $\Z\to\Pgt\Pgt$+jets backgrounds.

\subsubsection{Background modeling}
\label{sec:background_tautau}
The signal processes as well as irreducible background processes with the same lepton final state are modeled using simulated events.
These irreducible backgrounds include $\PW\Z$, $\Z\Z$, \ttH, and $\ttbar+\PW/\Z$ production.
The simulation is corrected for differences between data and simulation, including the distribution of pileup interactions, the efficiencies for the leptons to pass trigger, identification, and isolation criteria, and the identification efficiency for \cPqb{}~quark jets.

The contributions from reducible background processes are estimated using a similar tight-to-loose method as discussed in Section~\ref{sec:background_multileptons}.
The contributions from events where either the charge of an electron or muon is misreconstructed or the $\tauh$ candidate is misidentified are negligible.
Therefore, three control samples are defined where one or both leptons fail the tight identification and isolation criteria, but pass all other selections: (i) one of the leptons fails the tight criteria; (ii) the other lepton fails the tight criteria; and (iii) both leptons fail the tight criteria.

The tight-to-loose rates ($f$) for jets misidentified as electrons or muons are measured in control regions enriched in $\PW$+jets and $\ttbar$ events.
The selection criteria for these control regions differ from the signal selection by requiring the transverse mass of the leading isolated $\ell$-\MET system to be greater than 35\GeV and by requiring that there be no selected $\tauh$ leptons, making the selection orthogonal to the signal sample.
The rate $f$ is parameterized based on the lepton \pt and the number of jets with $\pt > 20$\GeV in the event, using the $k$-nearest neighbor algorithm~\cite{Hocker:2007ht}.
The small contributions from genuine isolated leptons from $\PW\Z$, $\Z\Z$, and $\ttbar+\Z/\PW$ events in the control region are estimated using simulated samples and subtracted.

For two leptons $\ell_1$ and $\ell_2$ with rates $f_1$ and $f_2$, the spectra of the reducible background contributions in the signal region are estimated by weighting events in the regions where only $\ell_1$ or $\ell_2$ fail the tight selection criteria by $f_1$ or $f_2$, respectively, and events in the region where both leptons fail the tight selection criteria by $- f_1 \, f_2$.
This procedure is conceptually identical to the one described in Section~\ref{sec:background_multileptons}.
The reducible background estimation is validated in a control region where the $\tauh$ candidates fail the tight isolation criteria.

\subsubsection{Signal extraction}
\label{sec:signal_extraction_tautau}

To extract the signal contribution, a multivariate method is used that combines the discrimination power of several variables.
The signal extraction is performed with a linear discriminant, also known as Fisher discriminant, as implemented in the TMVA package~\cite{Hocker:2007ht}.
Because of the small number of simulated and estimated background events in the signal region, the Fisher discriminant is trained using events from a control region with the \Pgt\ isolation criteria inverted.
This provides a sufficient number of training events and avoids overtraining from the events in the signal region, thereby improving the final expected sensitivity of the analysis.

The Fisher discriminant is trained using ten input variables making use of (i)~the forward jet present in \tHq\ production, (ii)~the expectation of only one \cPqb{}~quark jet as opposed to background processes including a $\ttbar$ pair, and (iii) other kinematic differences between the \tHq\ and the background processes.

The training variables are: $\abs{\eta}$ of the jet with the largest $\abs{\eta}$ value and $\pt > 20$\GeV, $\abs{\eta}$ of the jet with the largest $\abs{\eta}$ value and $\pt > 30$\GeV, ``centrality'' (the ratio of the \pt sum of all selected objects and the energy sum), number of b~jets, \pt of the leading b~jet, number of jets with $\pt > 30$\GeV, $\Pe\tauh$ invariant mass ($\Pgm\tauh$ mass with the leading muon in the \mmt channel), $\Pgm\tauh$ mass ($\Pgm\tauh$ mass with the subleading muon in the \mmt channel), $\Pe\Pgm$ mass ($\Pgm\Pgm$ mass in the \mmt channel), and \MET.
The training is performed assuming the \tHq process as a signal and the rest of the processes as background.
The tHW process is not considered as a part of the signal in the training because of its background-like shape,
but considered as a part of signal for the signal extraction.

\begin{figure}
 \includegraphics[width=0.45\textwidth]{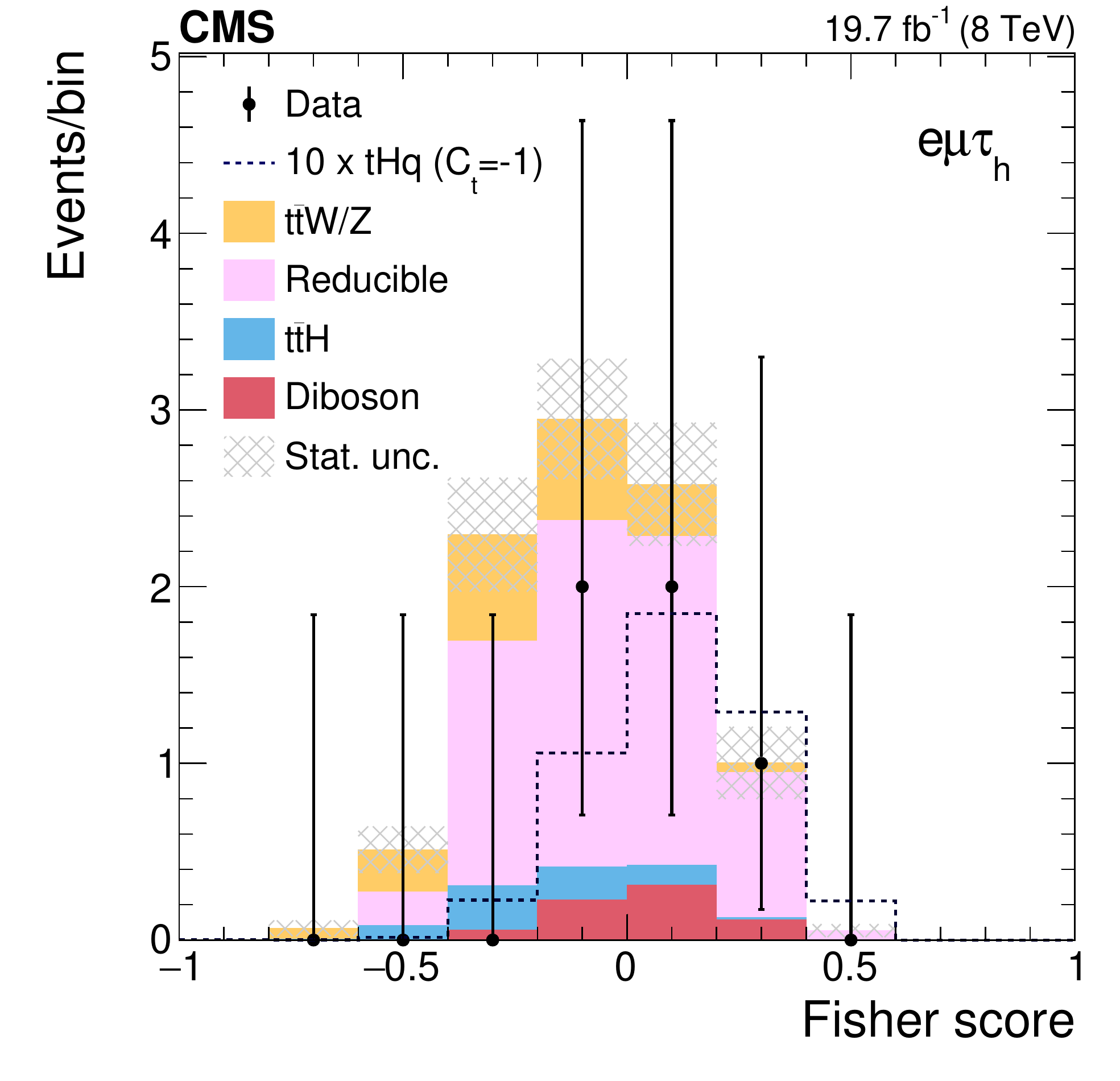}
 \includegraphics[width=0.45\textwidth]{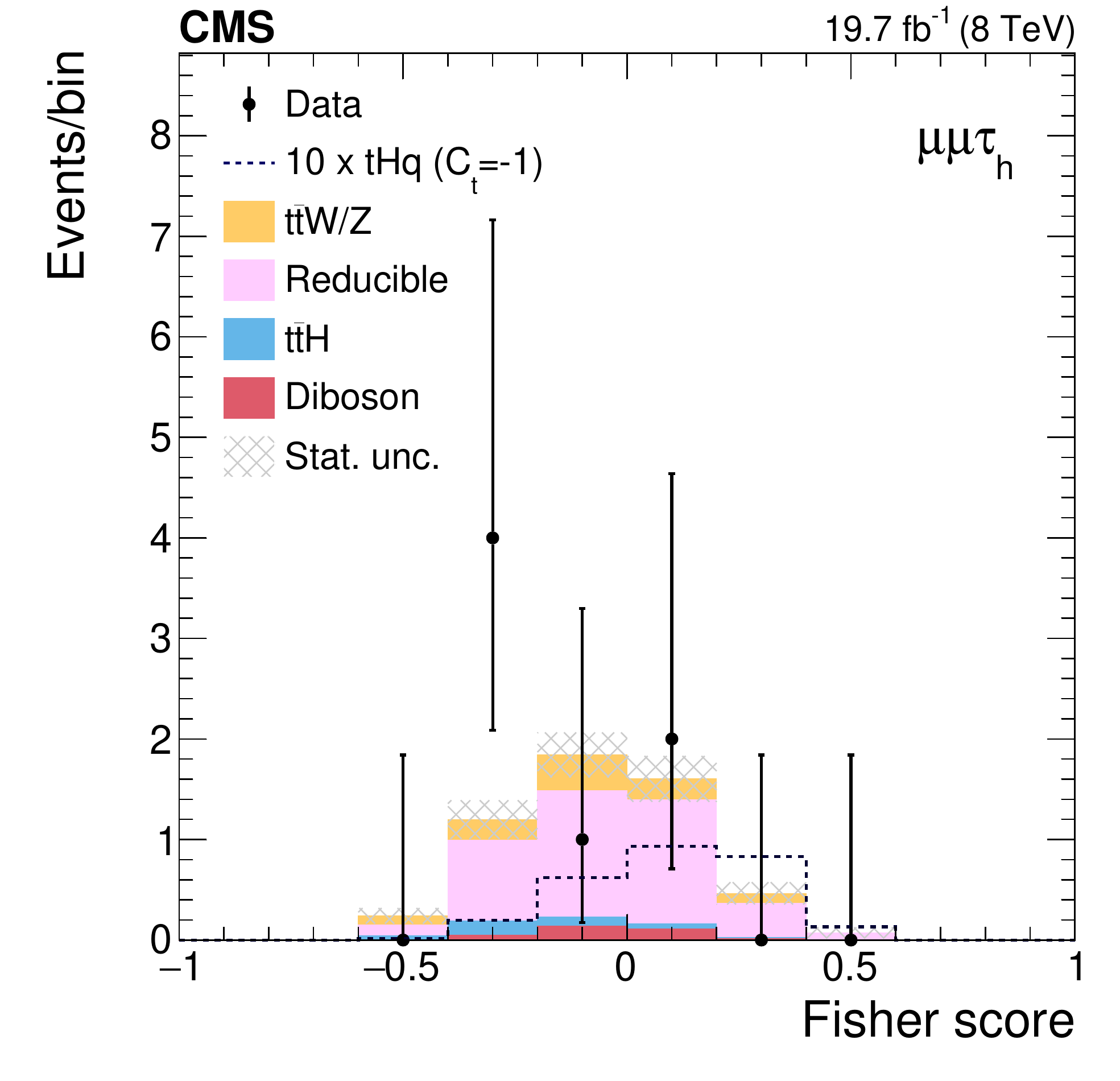}
  \caption{Expected (histograms) and observed (points) distributions of the Fisher discriminant in the $\Pe\mu\tauh$ channel (left) and $\mu\mu\tauh$ channel (right).
  The dashed line gives the expected contribution from the \tHq signal ($\Ct=-1$) case, multiplied by ten.}
  \label{fig:results_emt_mmt}
\end{figure}

The signal extraction is performed using a combined maximum likelihood fit of the Fisher discriminant distributions in the two channels.
Figure~\ref{fig:results_emt_mmt} shows the final distributions of the discriminant in the $\Pe\mu\tauh$ and $\mu\mu\tauh$ categories.
The expected and observed yields in all categories are given in Table~\ref{tab:final_yields}.

\begin{table}[htb]
  \topcaption{Expected and observed event yields for the $\Pe\mu\tauh$ and $\mu\mu\tauh$ channels. The given uncertainties include all systematic uncertainties added in quadrature, including uncertainties due to the limited numbers of simulated events or events in control data samples.}
  \centering
    \begin{tabular}{ccc}
      \hline
      Process          & \emt       & \mmt      \\
      \hline
      \tHq ($\Ct=-1$)   & 0.42 $\pm$ 0.05 & 0.26 $\pm$ 0.03 \\
      \tHW ($\Ct=-1$)   & 0.06 $\pm$ 0.01 & 0.04 $\pm$ 0.01 \\
      \hline
      \ttH             & 0.6  $\pm$ 0.1  & 0.3  $\pm$ 0.1  \\
      \ttV             & 1.8  $\pm$ 0.4  & 0.9  $\pm$ 0.2  \\
      VV               & 0.7  $\pm$ 0.1  & 0.3  $\pm$ 0.1  \\
      Reducible        & 6.3  $\pm$ 3.1  & 4.5  $\pm$ 1.9  \\
      \hline
      Total background & 9.5 $\pm$ 3.7  & 5.4  $\pm$ 2.4  \\
      \hline
      Data             & 5              & 7              \\
      \hline
    \end{tabular}
    \label{tab:final_yields}
\end{table}

\section{Systematic uncertainties}
\label{subsec:common_systs}

Various sources of systematic uncertainty influence the upper limit on the \tHq\ production cross section.
In general, the systematic uncertainties can introduce rate uncertainties on a specific process as well as shape uncertainties on the distribution from which the upper limit on the process is finally derived.
These  uncertainties are handled by means of nuisance parameters, which are allowed to float during the limit setting procedure.

The uncertainty in the trigger efficiencies translates into an uncertainty
in the final rates of up to 5\%.

The uncertainty from the jet energy scale~\cite{Chatrchyan:2011ds} is evaluated by varying the energy scale for all
jets in the signal and background simulation simultaneously within their
uncertainty as a function of jet $\pt$ and $\eta$, and re-evaluating the
yields and discriminant shapes of all processes.
The limitations on the knowledge of the jet energy scale lead to an uncertainty that in some channels can be as large as 8\%. Jet energy resolution uncertainties have a smaller effect, up to 3\% in the event yields.

The corrections for the $\cPqb$ tagging efficiencies for light-flavored, \cPqc, and \cPqb{}~quark jets have associated uncertainties~\cite{CMS-PAS-BTV-13-001}, which are parameterized as a function of the $\pt$, $\eta$, and flavor of the jets.
Their effect on the analysis is evaluated by shifting the correction factor of each jet up and down within their measured uncertainty.

For photon identification, the uncertainty in the data/MC efficiency scale factor from the fiducial region determines the overall uncertainty, as measured using a tag-and-probe technique applied to $\Z\to \Pe\Pe$ events (3.0\% in the ECAL barrel, 4.0\% in ECAL endcap)~\cite{Khachatryan:2010xn}.
For the uncertainties related to the photon energy scale and resolution, the photon energy is shifted and smeared, respectively, within the known uncertainty for photons~\cite{Chatrchyan:2013dga}.

The cross sections used to estimate signal and background rates, where applicable, are of at least NLO accuracy and have associated uncertainties arising primarily from the PDFs and the choice of the factorization and renormalization scales.

The effect from the PDF uncertainties has been evaluated on signal and
backgrounds following the PDF4LHC
prescription~\cite{Alekhin:2011sk,Botje:2011sn}, and ranges from 1 to 8\% depending  on the quark or gluon nature of the colliding partons.
The effect of changing renormalization and factorization scales is
evaluated for both signal and backgrounds by changing them simultaneously up and down by factors of two, producing effects on rates extending up to 13\% for \ttH\ production.
For the $\Hgg$ and $\HWW$ analyses, where the signal is modeled using the five-flavor scheme, the overall event selection efficiency is re-evaluated using a sample simulated with the four-flavor scheme.
The corresponding change in signal selection efficiency is taken as a systematic uncertainty, and is 5.5\% in the diphoton and up to 16\% in the multilepton channels.

The large \ttbar background in the $\PH \to \cPqb\cPaqb$ final state requires
a special treatment of the $\ttbar+\text{jets}$ background component, which is split into four different categories, depending on the flavor of the additional final state partons: $\ttbar+\cPqb\cPaqb$, $\ttbar+\PQb$, $\ttbar+1/2\PQc$, and $\ttbar+\text{light flavors}$. Each of the $\ttbar+\text{heavy flavor}$ components receives a conservative 50\% rate uncertainty in addition to what is assigned to the \ttbar background rate uncertainties.
Dedicated \MADGRAPH{}+\PYTHIA samples with varied renormalization and factorization scales and with varied matching thresholds are used to introduce additional nuisance parameters, which can alter the rate and the shape of the \ttbar backgrounds. Reweighting the top quark \pt distribution for \ttbar events needs to be accounted for by a separate rate and shape systematic uncertainty~\cite{Khachatryan:2015oqa}. The systematic uncertainty arising from scale variations in the sample generation is also taken into account for the signal process. For the statistical uncertainties, bin-by-bin uncertainties in the NN output shape are taken into account.

Uncertainties in the efficiencies for
lepton  identification, isolation and
impact parameter requirements are estimated by comparing variations in the
difference in performance between data and MC simulation using a high-purity
sample of \Z boson decays with a tag-and-probe method.
These uncertainties vary between 1 and 5\%, depending on the lepton flavor
and selection. The overall uncertainty is about 5\% per lepton for the same-sign dilepton final state, while it is 1.6\% in the case of the trilepton final state.
For trigger efficiencies, no scale factors are used on simulation to correct for possible differences between data and MC, assuming a trigger efficiency of 100\% for the double-lepton triggers.
The uncertainty in the yields derived from simulation due to the trigger efficiency is about 1\%.

The uncertainty in the misidentification probabilities for nonprompt leptons is estimated from simulation for the same-sign dilepton final state.
The misidentification rate is estimated following the same approach and parameterization used in the multijet-dominated control sample, but using instead MC samples with a similar composition.
This simulation-based misidentification rate is then applied to MC samples with the expected background composition in the signal sample, and the amount of disagreement between the number of nonprompt leptons predicted by the parameterized misidentification rate and those actually observed in this collection of MC samples is used to estimate the systematic uncertainty.
In the case of the same-sign dilepton final state, the uncertainty is assessed separately for different $\pt$, $\eta$, and \cPqb-tagged jet multiplicity bins for each flavor.
The overall uncertainty amounts to about 40\%, which is applied using linear and quadratic deformations of the $\pt$- and $\eta$-dependent misidentification rate.
For the trilepton final state, a similar method is used to estimate a total rate uncertainty of 30\%.
Additional sources of uncertainty for this final state are considered.
The first contribution comes from the change in the tight-to-loose rate as a result of applying a requirement on the \MET in the multijet control region used to estimate this rate, changing the diboson background contribution.
The overall effect on the final prediction is about 10\%.
The second contribution is studied by changing the measured tight-to-loose rate up and down within its statistical uncertainty and propagated to the final weight estimation.
The total effect on the expected number of events is about 14\%.

In the $\PH\to \tau\tau$ analysis, an uncertainty of 50\% is assigned to the yield of reducible backgrounds, uncorrelated between channels and categories.
This arises from the sum of the uncertainties in the estimation of the nonprompt rates with and without the requirement of a \cPqb-tagged jet, the agreement of the predicted and observed event yields in control regions, and the comparison with simulated \ttbar events.
An uncertainty of 6\% is assigned to the $\tauh$ identification efficiency~\cite{CMS-PAS-TAU-11-001}.
The $\tauh$ energy scale uncertainty is 3\%~\cite{Chatrchyan:2014nva}; this propagates to an uncertainty of comparable size in the simulated yields.

The uncertainty in the integrated luminosity, which is common to all of the channels, is 2.6\%~\cite{CMS-PAS-LUM-13-001}.
\section{Results}
\label{sec:res}
No significant excess of events over the expectations is observed in the different channels, and the observed yields and predicted backgrounds are used to set limits on the \tHq\ production cross section.
A binned likelihood spanning all analysis channels included in a given result is constructed.
Uncertainties in the signal and background predictions are incorporated by means of nuisance parameters.

Limits are computed using the modified-frequentist CL$_\mathrm{s}$ method~\cite{ALread:2012,Junk:1999kv}. Results are obtained independently for each of the distinct \tHq signatures (diphoton, $\cPqb\cPaqb$, WW, and $\tau\tau$)
as well as combined. There is no significant deviation in the data from the predicted event yields.

The $\Ct=-1$ scenario predicts an enhancement in the \tHq\ production cross section, and in the diphoton branching fraction of the Higgs boson. As a result, the presence of such a signal in the data would be highlighted in the diphoton channel by an enhancement of the yields in all Higgs boson production modes.

\begin{figure}[!htbp]
  {\centering
    \includegraphics[width=0.49\textwidth]{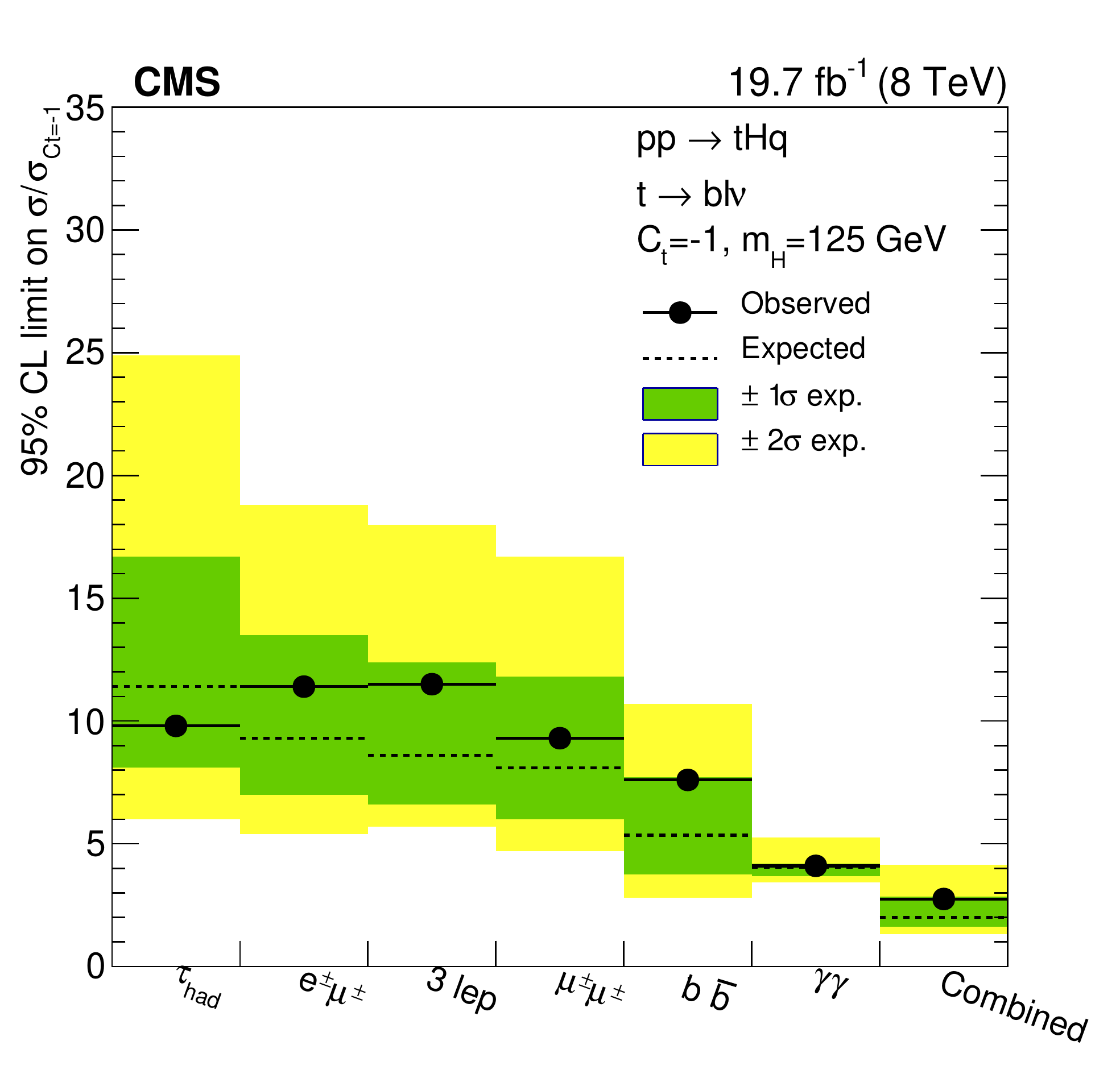}
  \caption{The 95\% CL upper limits on the excess event yields predicted by the enhanced \tHq\ cross section and Higgs boson to diphoton branching fraction for $\Ct=-1$. The limits are normalized to the $\Ct=-1$ predictions~\cite{MadGraph5_aMCNLO_2}, and are shown for each analysis channel, and combined. The black solid and dotted lines show the observed and background-only expected limits, respectively. The $1\sigma$ and $2\sigma$ bands represent the 1 and 2~standard deviation uncertainties on the expected limits.}
    \label{fig:Ct_limit}}
\end{figure}

The median expected 95\% confidence level (CL) upper limit computed from the combination of all channels is 2.0 times the event yields predicted by negative Yukawa couplings. The corresponding observed upper limit is 2.8. Table~\ref{tab:comb_limit_mu} shows the limits obtained in the several subchannels, and from the overall combination. Figure\,\ref{fig:Ct_limit} provides a visual display of the results.

\begin{figure}[!htbp]
  {\centering
    \includegraphics[width=0.49\textwidth]{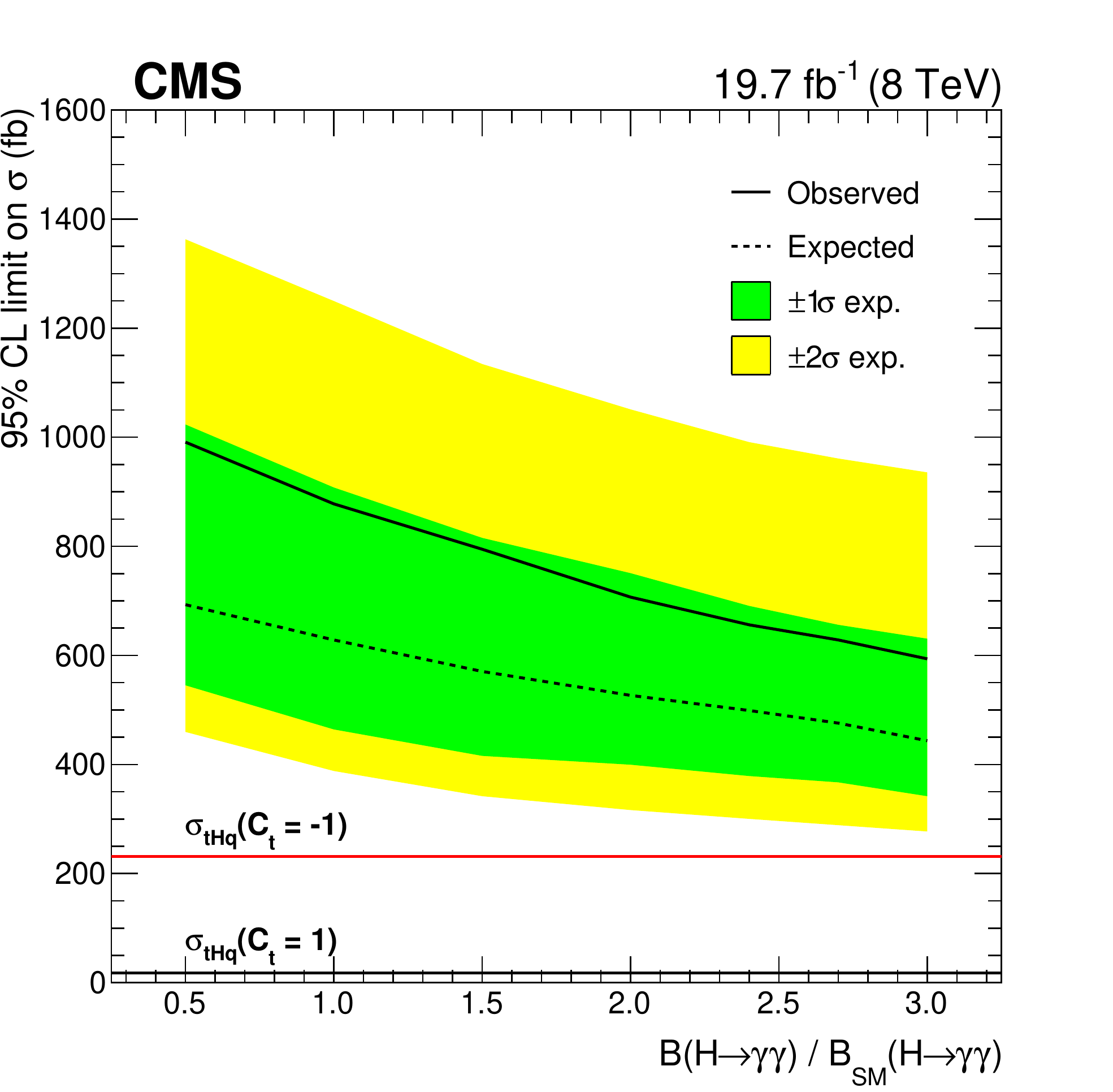}
    \caption{The 95\% CL upper limits on the \tHq\ production cross section as a function of the assumed Higgs boson to diphoton branching fraction. The black solid and dotted lines show the observed and background-only expected limits, respectively. The $1\sigma$ and $2\sigma$ bands represent the 1 and 2~standard deviation uncertainties on the expected limits.
The red horizontal line shows the predicted \tHq\ cross section for the SM Higgs boson with $m_{\PH}$ = 125\GeV\ in the $\Ct=-1$ scenario, while the black horizontal line shows the predicted \tHq\ cross section in the SM (\ie, $\Ct=+1$).}
    \label{fig:sigmaxBR_limit}}
\end{figure}

The 95\% CL expected and observed upper limits on the tHq production cross section are shown in Fig.~\ref{fig:sigmaxBR_limit} as a function of the assumed branching fraction for a Higgs boson decaying to two photons. The latter quantity is normalized to the SM predictions. In addition to the median expected limit under the background-only hypothesis, the bands that contain the one and two standard deviation ranges around the median are also shown.

The overall effect of systematic uncertainties on the analysis is to increase the upper limit for 95\% exclusion by $\sim$40\%, averaged over the range of BR(H $\rightarrow \gamma \gamma$) values investigated.

\begin{table}[!htbp]
  \centering
\topcaption{Upper limit on $\mu = \sigma/\sigma_{\Ct=-1}$ for each \tHq channel. The observed and expected 95\% CL upper limits on the signal strength parameter $\mu$ for each \tHq channel are also shown. }
    \begin{tabular}{ccccc}
      \hline
      \tHq channel & Best-fit $\mu$ &  \multicolumn{3}{c}{95\% CL upper limits on $\mu = \sigma/\sigma_{\Ct=-1}$}
      \\ \hline
      &  Observed &  \multicolumn{3}{c}{Expected} \\
      &           &   \multirow{2}{*}{Median}  &  \multirow{2}{*}{68\% CL range}  &  \multirow{2}{*}{95\% CL range}  \\
      \\\rule[-1.4ex]{0pt}{4ex} $\gamma\gamma$ & 4.1 & 4.1  & [3.7, 4.2]  & [3.4, 5.3] \\
      \rule[-1.4ex]{0pt}{4ex} $\bbbar$  & 7.6 & 5.4  & [3.8, 7.7]  & [2.8, 10.7] \\
      \rule[-1.4ex]{0pt}{4ex} Multilepton      & 6.7 & 5.0  & [3.6, 7.1]  & [2.9, 10.3] \\
      \rule[-1.4ex]{0pt}{4ex} $\tau\tau$       & 9.8 & 11.4 & [8.1, 16.7] & [6.0, 24.9] \\ \hline
      \rule[-1.4ex]{0pt}{4ex} Combined         & 2.8 & 2.0  & [1.6, 2.8]  & [1.2, 4.1] \\
      \hline
    \end{tabular}
  \label{tab:comb_limit_mu}
\end{table}

Expected limits on the \tHq\ production cross section are set, ranging from 450 to 700\unit{fb} depending on the assumed diphoton branching fraction of the Higgs boson. The observed limits are slightly less sensitive, ranging from 600 to 1000\unit{fb}.
\section{Summary}
\label{sec:summ}

The production of the standard-model-like Higgs boson in association with a single top quark has been investigated using data recorded by the CMS experiment at $\sqrt{s}=8\TeV$, corresponding to an integrated luminosity of 19.7\fbinv.
Signatures resulting from leptonic top quark decay and different decay modes
of the Higgs boson have been analyzed.
In particular, the searches have been optimized for the $\PH \to \gamma \gamma$, $\PH \to \cPqb\cPaqb$, $\PH \to \PW\PW$, and $\PH \to \Pgt\Pgt$ decay modes.
The results are consistent with the background-only hypothesis.
A 95\% confidence level limit on the production cross section of a single top quark plus a Higgs boson with a non-standard-model coupling is set ranging from 600 to 1000\unit{fb} depending on the assumed diphoton branching fraction of the Higgs boson. This is the first time that results on anomalous \tHq\ production have been reported. These results can be combined with other Higgs boson measurements to constrain the coupling of the Higgs boson to SM quarks; they can also be used to probe new physics modifying the top-Higgs couplings. The 13\TeV LHC run will allow a precise determination of both the magnitude and the sign of the top quark Yukawa coupling.

\begin{acknowledgments}
We congratulate our colleagues in the CERN accelerator departments for the excellent performance of the LHC and thank the technical and administrative staffs at CERN and at other CMS institutes for their contributions to the success of the CMS effort. In addition, we gratefully acknowledge the computing centers and personnel of the Worldwide LHC Computing Grid for delivering so effectively the computing infrastructure essential to our analyses. Finally, we acknowledge the enduring support for the construction and operation of the LHC and the CMS detector provided by the following funding agencies: BMWFW and FWF (Austria); FNRS and FWO (Belgium); CNPq, CAPES, FAPERJ, and FAPESP (Brazil); MES (Bulgaria); CERN; CAS, MoST, and NSFC (China); COLCIENCIAS (Colombia); MSES and CSF (Croatia); RPF (Cyprus); MoER, ERC IUT and ERDF (Estonia); Academy of Finland, MEC, and HIP (Finland); CEA and CNRS/IN2P3 (France); BMBF, DFG, and HGF (Germany); GSRT (Greece); OTKA and NIH (Hungary); DAE and DST (India); IPM (Iran); SFI (Ireland); INFN (Italy); MSIP and NRF (Republic of Korea); LAS (Lithuania); MOE and UM (Malaysia); CINVESTAV, CONACYT, SEP, and UASLP-FAI (Mexico); MBIE (New Zealand); PAEC (Pakistan); MSHE and NSC (Poland); FCT (Portugal); JINR (Dubna); MON, RosAtom, RAS and RFBR (Russia); MESTD (Serbia); SEIDI and CPAN (Spain); Swiss Funding Agencies (Switzerland); MST (Taipei); ThEPCenter, IPST, STAR and NSTDA (Thailand); TUBITAK and TAEK (Turkey); NASU and SFFR (Ukraine); STFC (United Kingdom); DOE and NSF (USA).

Individuals have received support from the Marie-Curie program and the European Research Council and EPLANET (European Union); the Leventis Foundation; the A. P. Sloan Foundation; the Alexander von Humboldt Foundation; the Belgian Federal Science Policy Office; the Fonds pour la Formation \`a la Recherche dans l'Industrie et dans l'Agriculture (FRIA-Belgium); the Agentschap voor Innovatie door Wetenschap en Technologie (IWT-Belgium); the Ministry of Education, Youth and Sports (MEYS) of the Czech Republic; the Council of Science and Industrial Research, India; the HOMING PLUS program of the Foundation for Polish Science, cofinanced from European Union, Regional Development Fund; the OPUS program of the National Science Center (Poland); the Compagnia di San Paolo (Torino); the Consorzio per la Fisica (Trieste); MIUR project 20108T4XTM (Italy); the Thalis and Aristeia programs cofinanced by EU-ESF and the Greek NSRF; the National Priorities Research Program by Qatar National Research Fund; the Rachadapisek Sompot Fund for Postdoctoral Fellowship, Chulalongkorn University (Thailand); and the Welch Foundation, contract C-1845.
\end{acknowledgments}

\bibliography{auto_generated}

\clearpage
\numberwithin{figure}{section}
\appendix

\section{Selected distributions of inputs to the multivariate discriminants}
The cross section for single top plus Higgs production in proton-proton collisions at $\sqrt{s} = 8$\TeV is very small, even under the hypothesis of an anomalous Yukawa coupling that is explored in this paper. In addition, the kinematics of the \tHq process are only mildly different from the dominant background processes.  Multivariate techniques are thus used to improve signal-to-background discrimination. Such techniques are used either at event selection level (in the case of the \Hgg analysis) or at signal discrimination level (in the case of \HWW and $\PH\to \tau\tau$) or in multiple steps of the analysis (in the case of \Hbb).  A representative selection of the kinematic distributions that are inputs to the multivariate discriminants are given in this Appendix.

\begin{figure}[th!]
        \centering
        \includegraphics[width=0.4\textwidth]{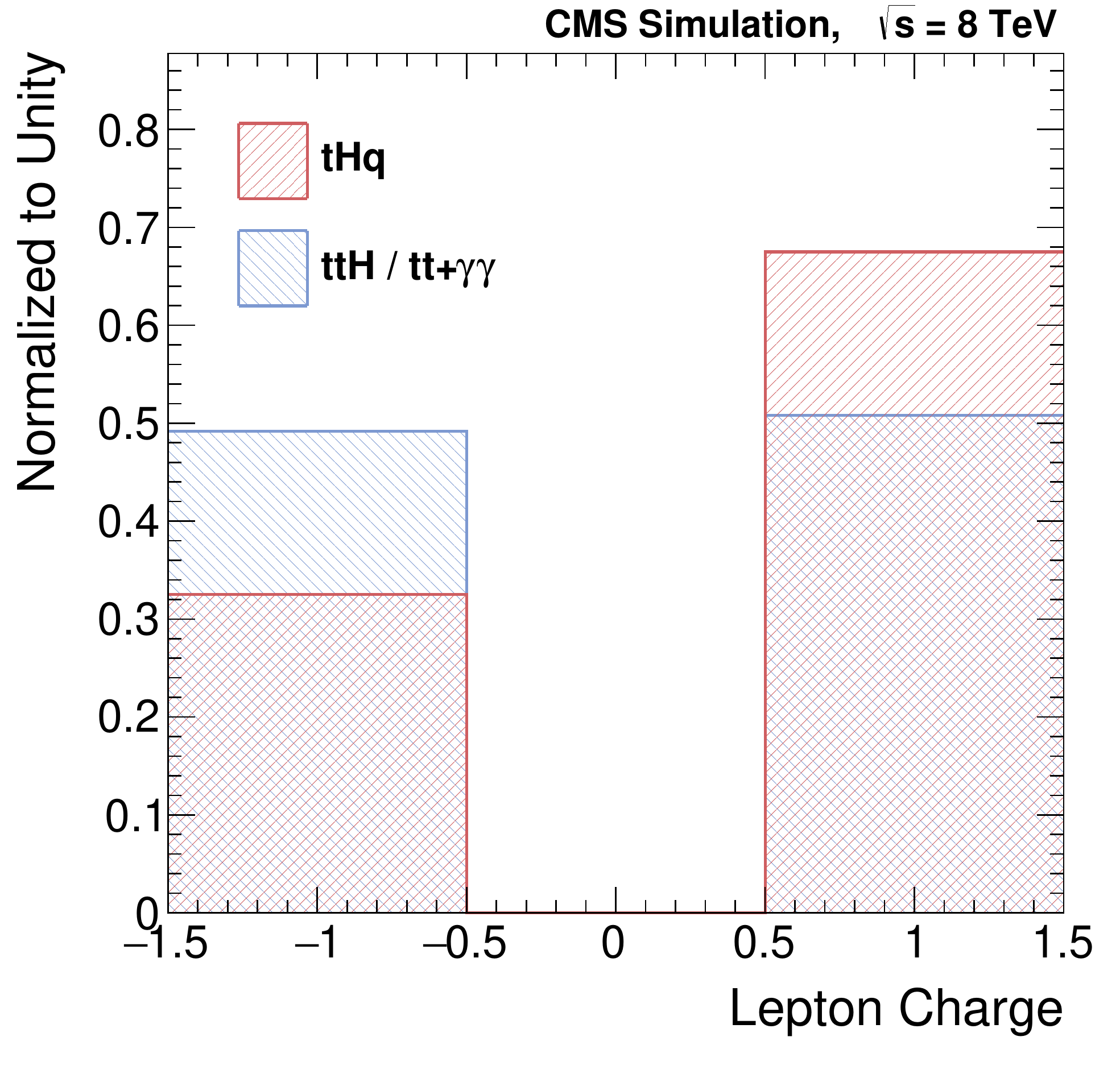}
        \includegraphics[width=0.4\textwidth]{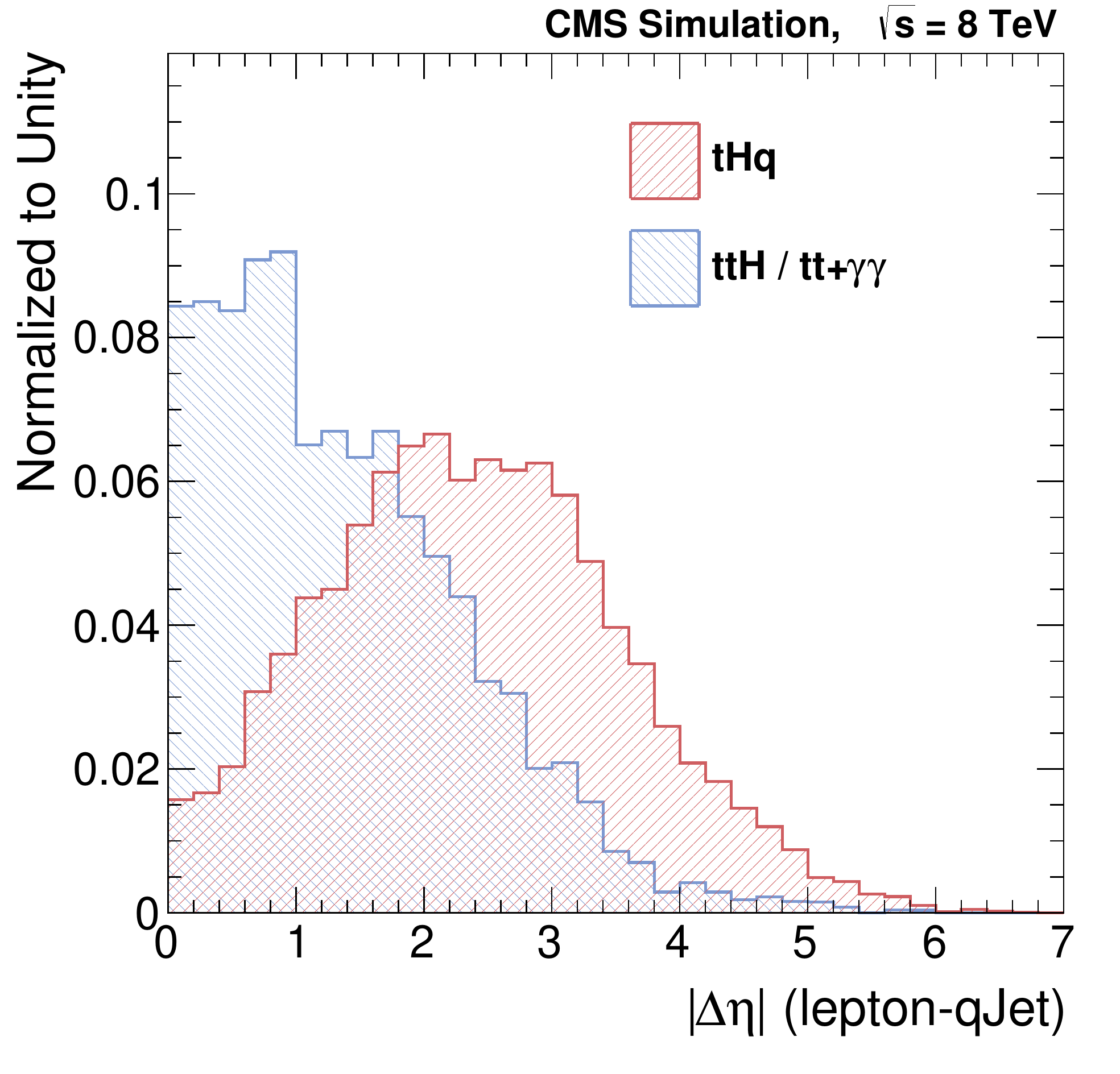}
        \caption{Sample of input variables to the final discriminant in the \Hgg
 analysis. Left: distribution of the electric charge of the reconstructed lepton. Right: distribution of the pseudorapidity difference between the lepton and the forward jet. Both plots are shown after the initial kinematic selection, and both distributions are normalized to unit area. \label{fig:app_diph}}
\end{figure}

One feature common to all analyses is the presence of a charge asymmetry induced by the net positive charge of the proton-proton collisions. This is displayed in the context of the \Hgg analysis in Fig.\,\ref{fig:app_diph}, together with the difference in pseudorapidity between the light-flavor jet and the charged lepton.

\begin{figure}[th!]
        \centering
        \includegraphics[width=0.45\textwidth]{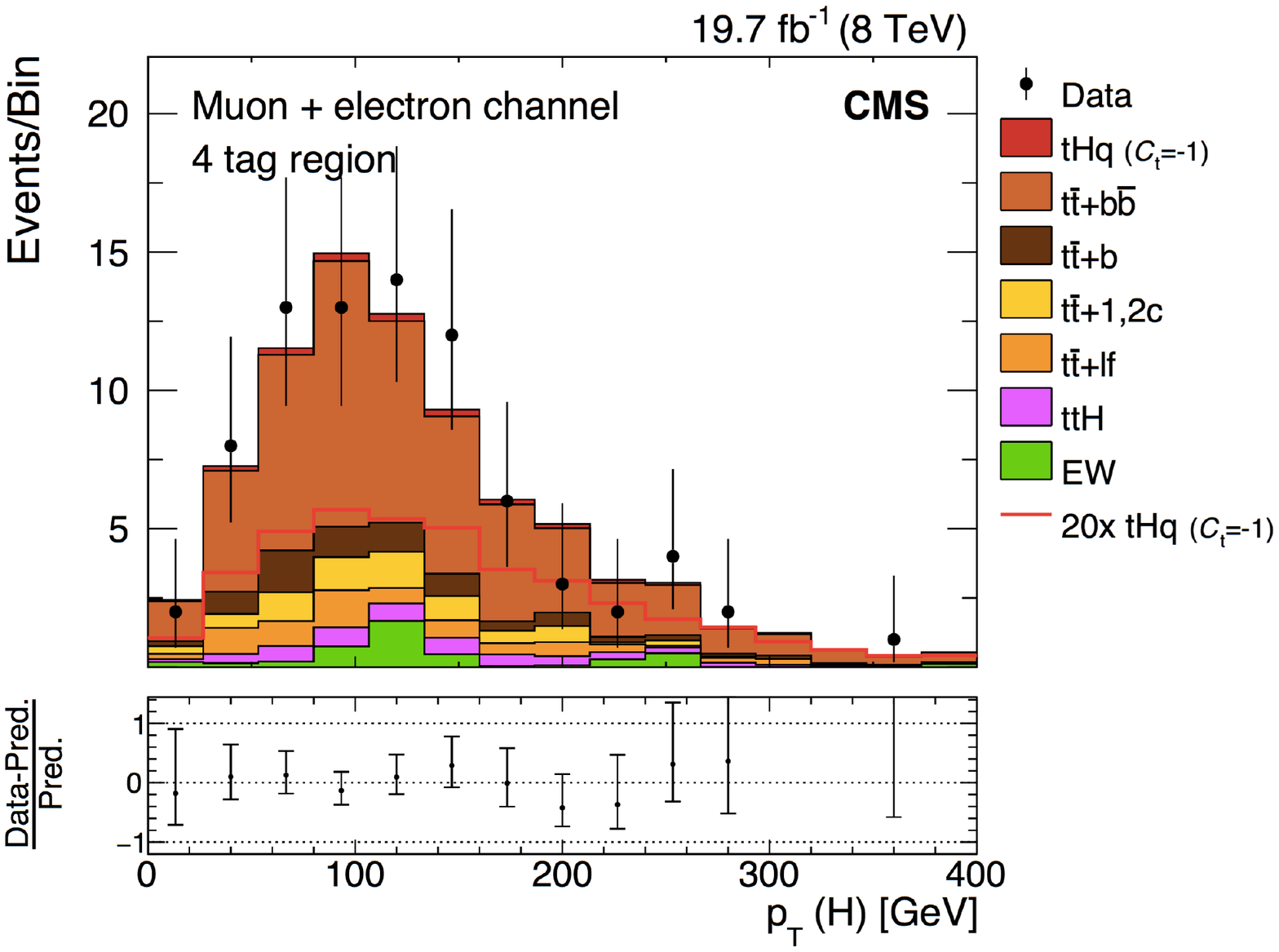}
        \includegraphics[width=0.45\textwidth]{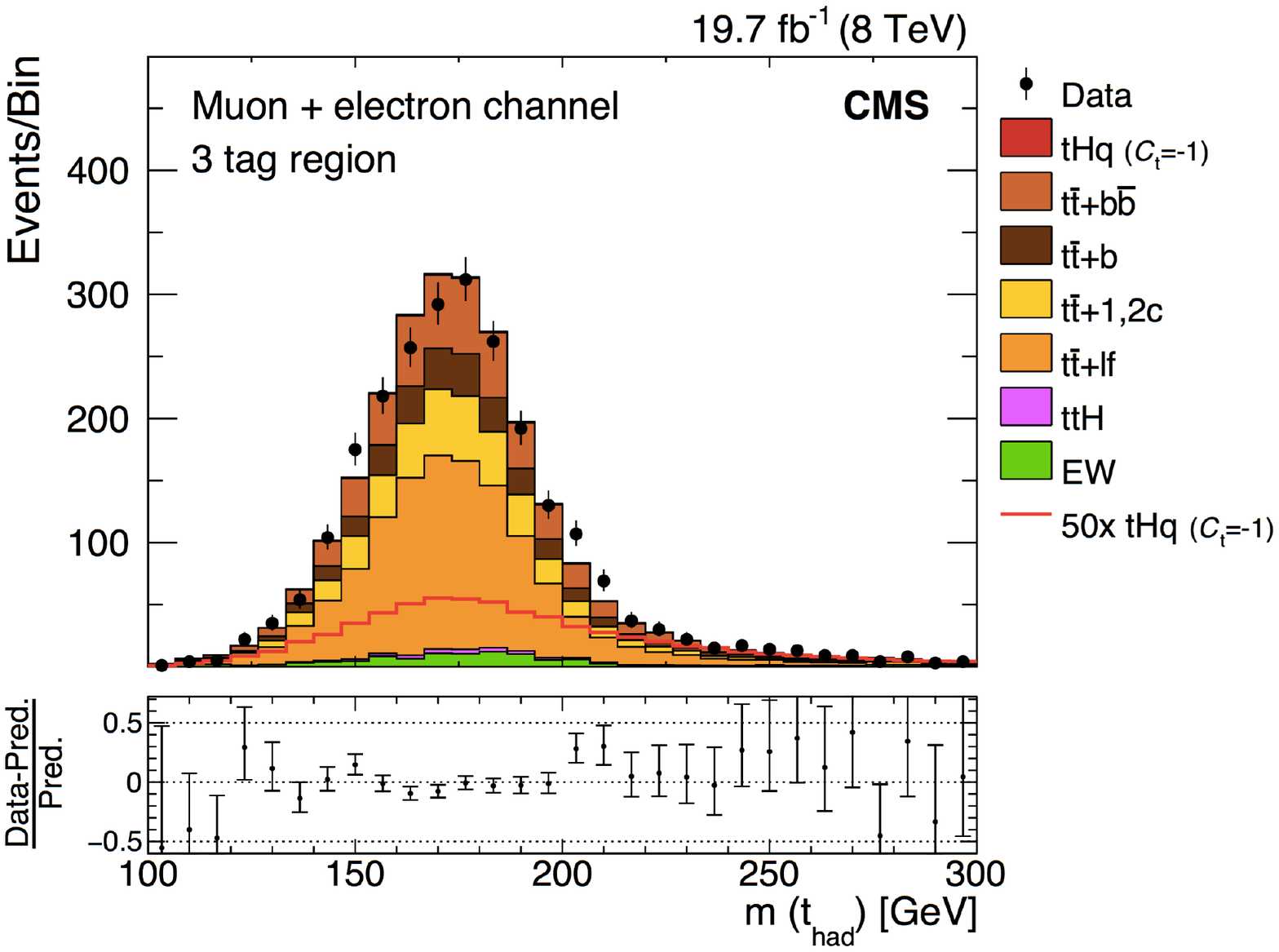}
        \caption{Sample of input variables to the final discriminant in the \Hbb channel. Left: distribution of the reconstructed transverse momentum of the $\bbbar$ system in the signal hypothesis for events with an high transverse momentum electron or muon and four b-tagged jets. Right: distribution of the invariant mass of the tri-jet system under the hypothesis of the presence of a hadronically-decaying top quark for electron/muon events with exactly three b-tagged jets. In both cases the signal is normalized to the rate for the analous coupling hypothesis, and is shown stacked on top of the SM background prediction.  It is also shown as a separate histogram, magnified by a factor of 20 (50) in the four-tag (three-tag) sample to enhance visibility. \label{fig:app_bbar}}
\end{figure}

In the \Hbb analysis, full event reconstructions are performed under both the \tHq and the \ttbar hypotheses. Fig.\,\ref{fig:app_bbar} shows the invariant mass of the tri-jet system most compatible with the hypothesis of a hadronically-decaying top quark and the transverse momentum of the b-jet pair compatible with originating from the decay of a Higgs boson. Good agreement between data and predictions is observed.

\begin{figure}[th!]
        \centering
        \includegraphics[width=0.37\textwidth]{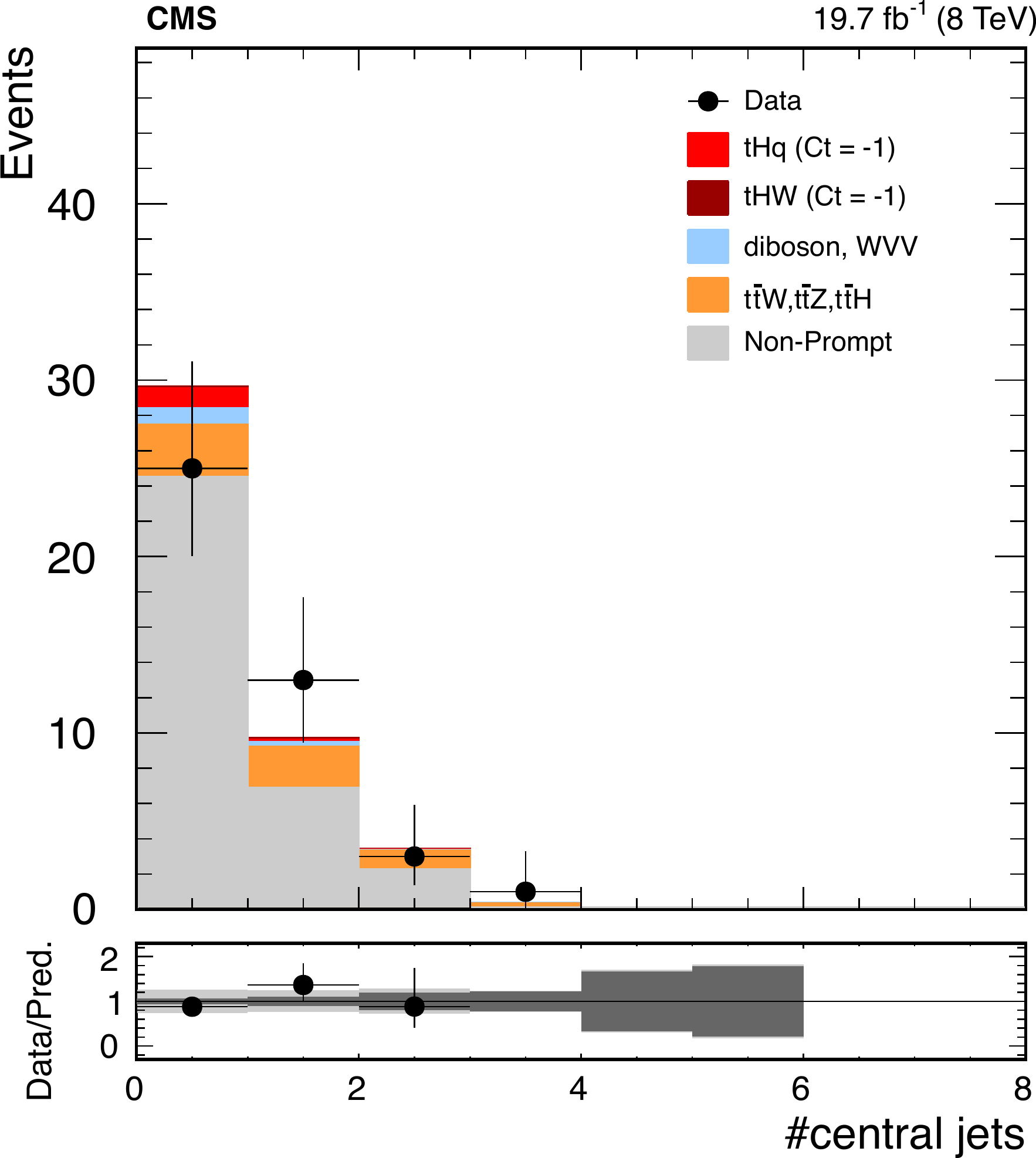}
        \includegraphics[width=0.37\textwidth]{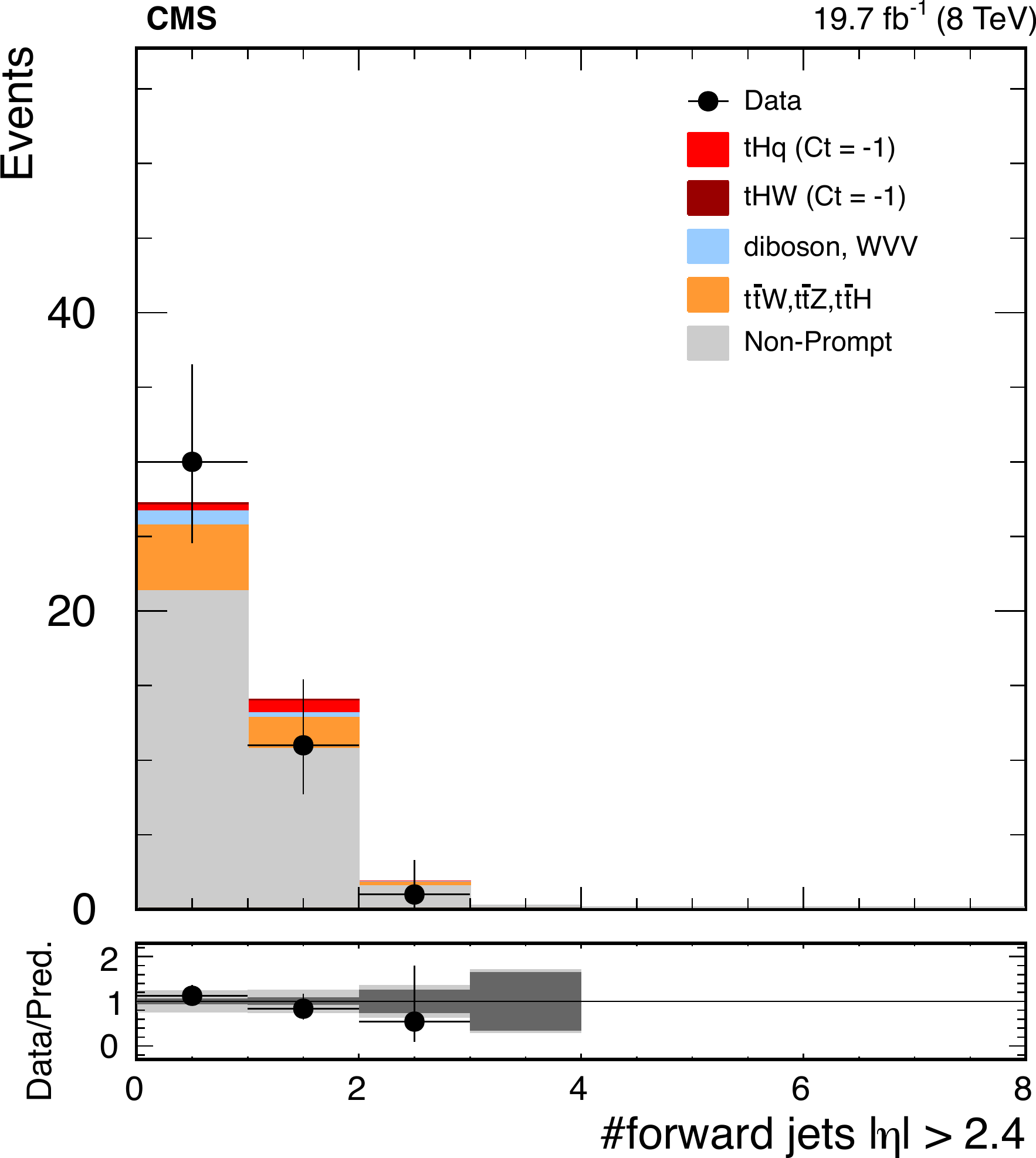}
        \caption{Sample of input variables to the final discriminant in the \HWW analysis. Left: distribution of the number of central jets. Left: distribution of the number of forward jets. The signal contribution is normalized to the prediction of the anomalous coupling hypothesis and stacked on top of the predicted background contributions. The bottom panels show the ratio between data and predictions, together with the statistical and systematic uncertainty bands. \label{fig:app_lep}}
\end{figure}

In the \HWW analysis, the event reconstruction is complicated by the presence of multiple neutrinos. The multiplicity of central jets ($\abs{\eta} < 1$) and forward jets ($\abs{\eta} > 1.5$) discriminates well between the \tHq signal and the background from non-prompt leptons, which originates mostly from \ttbar production. Both distributions are shown in Fig.\,\ref{fig:app_lep} for the events containing three charged leptons.

\begin{figure}[th!]
        \centering
        \includegraphics[width=0.4\textwidth]{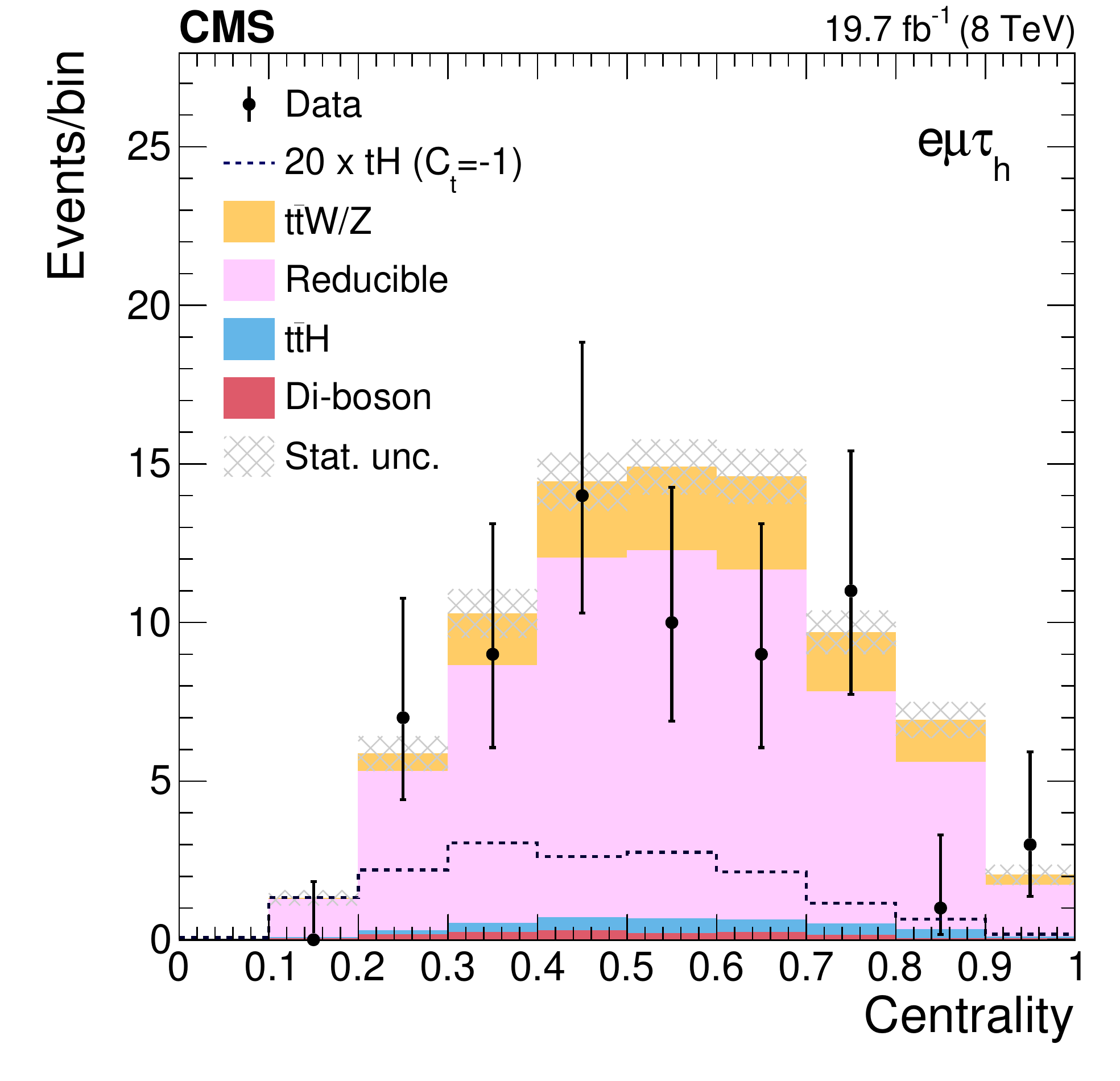}
        \includegraphics[width=0.4\textwidth]{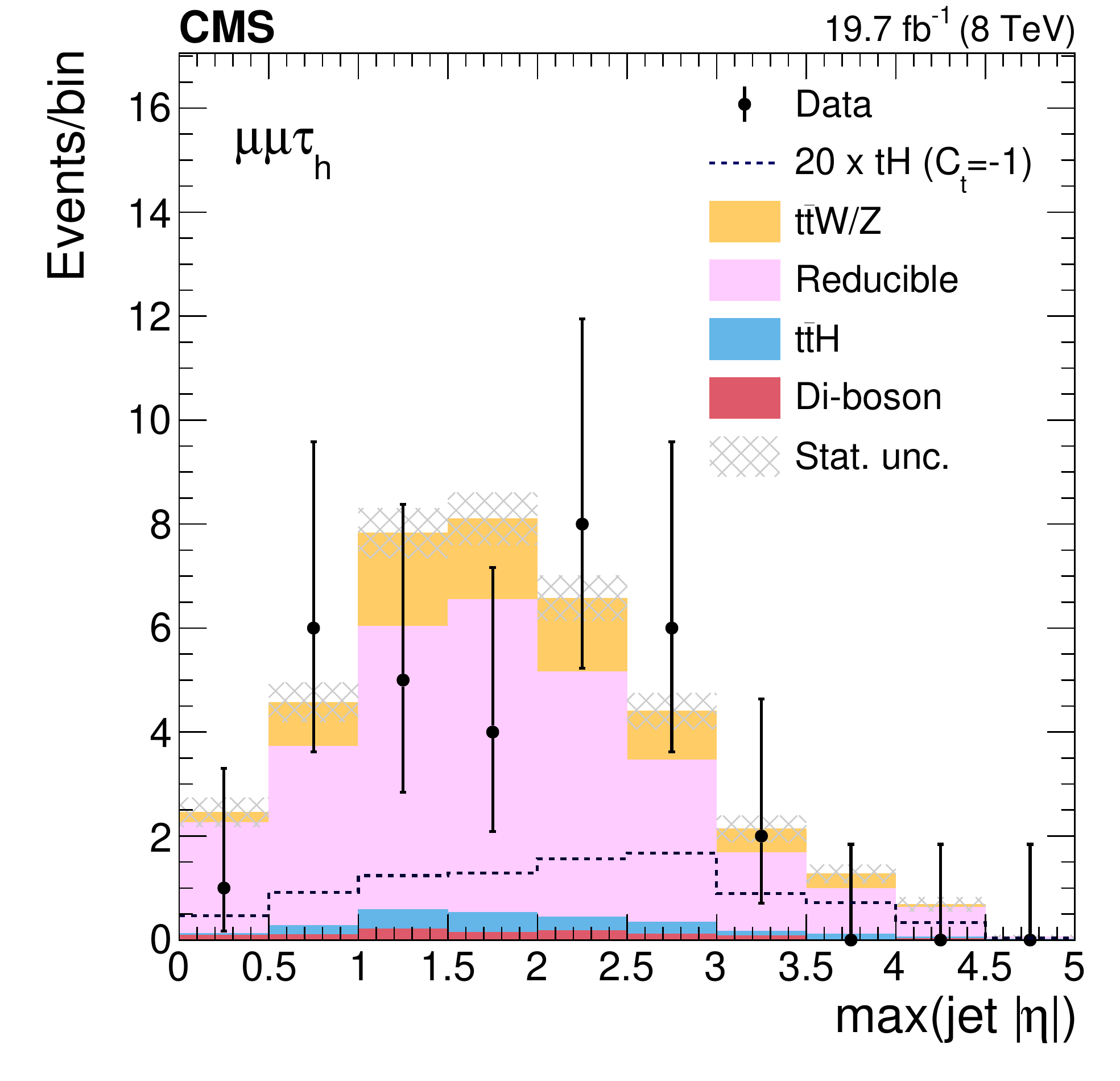}
        \caption{Sample of input variables to the final discriminant in the $\PH\to \tau\tau$ analysis. The distributions are shown for events failing the tight electron and muon isolation criteria. Left: distribution of the centrality for events with an electron, a muon, and a hadronically-decaying tau lepton. Right: distribution of the rapidity of the most forward jet in events with two muons and a hadronically-decaying tau lepton. The dashed line shows the expected contribution from the \tHq signal ($\Ct = - 1$) case, multiplied by 20. \label{fig:app_tau}}
\end{figure}

In the $\PH\to \tau\tau$ analysis it is important to understand both the kinematics and the rate of the reducible backgrounds. The background estimation from leptons falsely identified as hadronically-decaying $\tau$ leptons is validated in a control region where the light-lepton same-charge requirement is maintained, but the isolation requirement on the \tauh candidate is inverted. This region is dominated by W+jets and \ttbar+jets backgrounds. The agreement between data and predictions in this control region is shown for the \emt and \mmt channels in Fig.\,\ref{fig:app_tau}. The residual differences are taken into account as systematic uncertainties in the analysis.

\cleardoublepage \section{The CMS Collaboration \label{app:collab}}\begin{sloppypar}\hyphenpenalty=5000\widowpenalty=500\clubpenalty=5000\input{HIG-14-027-authorlist.tex}\end{sloppypar}
\end{document}

%% file: HIG-14-027-authorlist.tex
\textbf{Yerevan Physics Institute,  Yerevan,  Armenia}\\*[0pt]
V.~Khachatryan, A.M.~Sirunyan, A.~Tumasyan
\vskip\cmsinstskip
\textbf{Institut f\"{u}r Hochenergiephysik der OeAW,  Wien,  Austria}\\*[0pt]
W.~Adam, E.~Asilar, T.~Bergauer, J.~Brandstetter, E.~Brondolin, M.~Dragicevic, J.~Er\"{o}, M.~Flechl, M.~Friedl, R.~Fr\"{u}hwirth\cmsAuthorMark{1}, V.M.~Ghete, C.~Hartl, N.~H\"{o}rmann, J.~Hrubec, M.~Jeitler\cmsAuthorMark{1}, V.~Kn\"{u}nz, A.~K\"{o}nig, M.~Krammer\cmsAuthorMark{1}, I.~Kr\"{a}tschmer, D.~Liko, T.~Matsushita, I.~Mikulec, D.~Rabady\cmsAuthorMark{2}, B.~Rahbaran, H.~Rohringer, J.~Schieck\cmsAuthorMark{1}, R.~Sch\"{o}fbeck, J.~Strauss, W.~Treberer-Treberspurg, W.~Waltenberger, C.-E.~Wulz\cmsAuthorMark{1}
% ...
\vskip\cmsinstskip
% ...
1:~~Also at Vienna University of Technology, Vienna, Austria\\
2:~~Also at CERN, European Organization for Nuclear Research, Geneva, Switzerland\\
% ...
